\documentclass[11pt, a4paper]{article}
\usepackage{jheppub}
\usepackage{bm} 
\usepackage{subcaption}

\newtheorem{thm}{Theorem}

\def\bea{\begin{eqnarray}}
\def\eea{\end{eqnarray}}
\def\beas{\begin{eqnarray*}}
\def\eeas{\end{eqnarray*}}

\title{All-loop cuts from the Amplituhedron}
\author[a]{Cameron Langer,}
\author[b]{Akshay Yelleshpur Srikant,}
\affiliation[a]{Center for Quantum Mathematics and Physics (QMAP), University of California, Davis, CA, USA}
\affiliation[b]{Department of Physics, Princeton University, NJ, USA}
\abstract{The definition of the amplituhedron in terms of sign flips involves both one-loop constraints and the ``mutual positivity" constraint. To gain an understanding of the all-loop integrand of $\mathcal{N}=4$ sYM requires understanding the crucial role played by mutual positivity. This paper is an attempt towards developing a procedure to introduce the complexities of mutual positivity in a systematic and controlled manner. As the first step in this procedure, we trivialize these constraints and understand the geometry underlying the remaining constraints to all loops and multiplicities. We present a host of configurations which correspond to various faces of the amplituhedron. The results we derive are valid at all multiplicities and loop orders for the maximally helicity violating (MHV) configurations. These include detailed derivations for the results in \cite{Arkani-Hamed:2018caj}. We conclude by indicating how one might move beyond trivial mutual positivity by presenting a series of  configuration which re-introduce it bit by bit.}
\begin{document}
\maketitle
\newpage
\section{Introduction}
The amplituhedron is a geometric object that is conjectured to encode all the perturbative scattering amplitudes of planar $\mathcal{N}=4$ sYM. First introduced in \cite{TheAmplituhedron}, the original definition of this object was built on the discovery of the structures of the positive Grassmannian uncovered in \cite{positiveGrass} as well as the observation in \cite{hodges} associating the NMHV tree amplitude to the volume of a particular polytope in momentum twistor space. The amplituhedron realizes a similar geometric picture for general tree amplitudes and loop integrands, associating to each positive geometry a (conjecturally unique) ``canonical differential form'' defined by having logarithmic singularities on all its boundaries \cite{Arkani-Hamed:2017tmz}. The computation of scattering amplitudes in planar $\mathcal{N}=4$ is equivalent to determining a triangulation of the amplituhedron, so that different representations of amplitudes correspond to different geometric triangulations of the space. There is nontrivial evidence \cite{Bern:2014kca,Bern:2015ple,Bern:2018oao} that this geometric construction can be extended to the nonplanar sector of the theory, as the essential analytic properties of the loop integrand, namely logarithmic singularities and no poles at infinity \cite{Arkani-Hamed:2014via}, have been observed to hold beyond the planar limit.

 Understanding this geometry $\mathcal{A}_{n,k,L}$ for all multiplicities $n$, helicity configurations $k$ and loop orders $L$ is an open problem, and many different directions have been explored. The connections between the tree level amplituhedron and the Yangian symmetry of $\mathcal{N}=4$ have been explored in \cite{Ferro:2016zmx}, while a triangulation-independent understanding of the geometry has been studied from several different perspectives \cite{Ferro:2018vpf,Enciso:2014cta,Enciso:2016cif}, primarily for NMHV trees. An explicit description of how the BCFW cells triangulate the tree-level space was given in \cite{Karp:2017ouj} while an alternative sign flip reformulation of the $m=1$ amplituhedron was given in \cite{Karp:2016uax}. A manifestly Yangian invariant diagrammatic formulation using so-called ``momentum twistor diagrams'' was introduced in \cite{Bai:2014cna} and used to study the structure of the one-loop geometry in \cite{Bai:2015qoa}.  The higher loop-level geometry of the amplituhedron was explored in detail in \cite{IntoTheAmplituhedron,Anatomy} and an attempt to completely understand the geometry at four points and progressively higher loops can be found in \cite{4pt1, 4pt2, 4pt3}. However, important open questions regarding the technical details of triangulating the amplituhedron remain. Moreover, while the original definition provided a deeper understanding of the positive Grassmannian and on-shell diagrammatic structure of scattering amplitudes in $\mathcal{N}=4$ sYM, it was still slightly unsatisfactory since all these structures were associated to an auxiliary space not directly tied to the kinematic data.
 
 The introduction of the topological definition of the amplituhedron in \cite{Arkani-Hamed:2017vfh} completely resolved this issue, revealing the geometric structure of the amplitudes directly in kinematic space. In this new formulation, the amplitudes and loop integrands could now be thought of as differential forms in momentum twistor space depending on the loop integration variables as well as the external data. Recently, it was discovered that the scattering amplitudes in other theories may also be written as differential forms on the space of kinematical data, see e.g, \cite{notesdiff, associahedron, halohedron1}.

The topological definition also makes it clear that the inequalities that define the multi-loop amplituhedron fall into two categories. The first set of conditions constrains the variable associated with each loop to live in the one-loop amplituhedron, while the second set of conditions enforces mutual positivity among the different loops. This division provides us with greater control on the source of complexity -- the mutual positivity. A full understanding of the interplay between these two conditions is still lacking. However, as a starting point we begin by analyzing special configurations which completely trivialize mutual positivity. These cuts are exactly the opposite of the all-loop cuts considered in \cite{IntoTheAmplituhedron}, which focus on cutting propagators involving external data. Moreover, we begin an investigation of the effects of mutual positivity by introducing this non-triviality in stages. 

\begin{figure}[t!]
\begin{center}
\includegraphics[scale=.5]{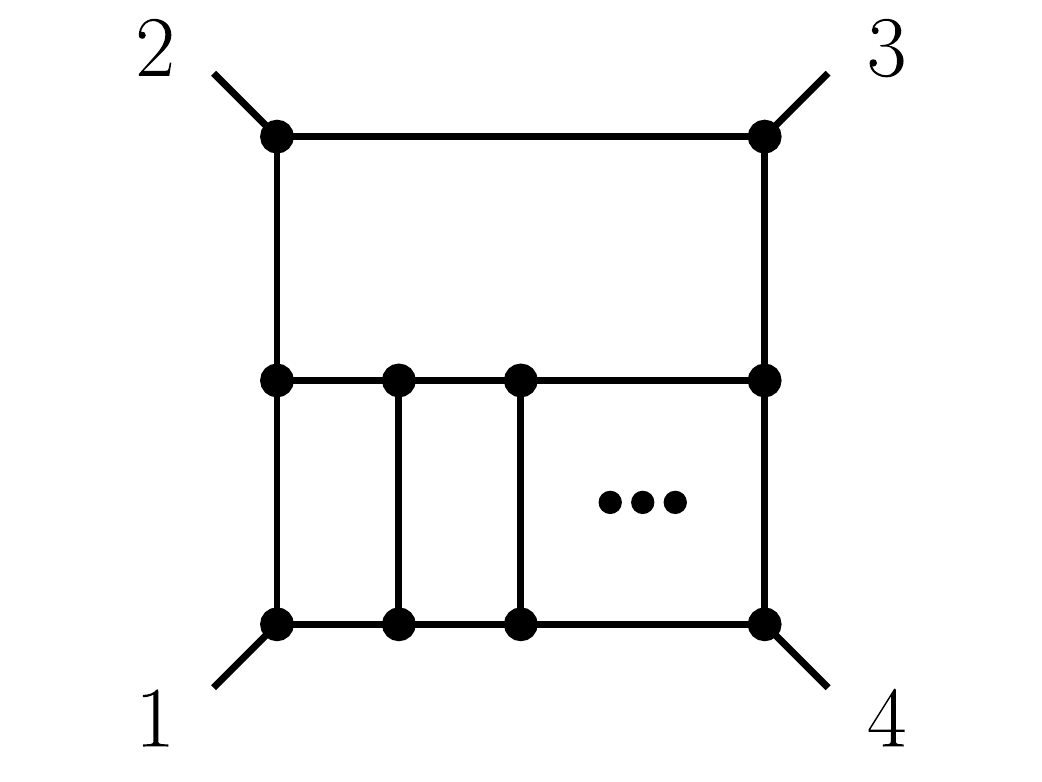}
\end{center}
\caption{A subclass of diagrams which contribute to the cuts considered in this paper.}
\label{fig:diagrams}
\end{figure}

One way to understand the geometry of the all-loop amplituhedron is by exploring different cuts of the loop integrand. In addition to specifying the structure of the amplituhedron's boundaries, these cuts allow us to access all-loop order information about the loop integrand which seems out of reach using any other known method. In this paper, we utilize the reformulation of the amplituhedron outlined in \cite{Arkani-Hamed:2017vfh} to explore a few faces of the all-loop MHV amplituhedron. These will involve cutting the maximal number of internal propagators involving loop momenta and thus trivializing the mutual positivity conditions between loops. As an example, in terms of Feynman diagrams, at four points, our cut will include (but is not limited to) summing over all diagrams of the form shown in Figure~\ref{fig:diagrams}.
In this sense the cuts we consider in this work probe the contributions of the most complicated multi-loop Feynman diagrams to the loop integrand involving the highest number of internal propagators. We will derive compact expressions for these cuts which are valid at all loop orders and, moreover, for an arbitrary number of external particles. This is a companion paper to \cite{Arkani-Hamed:2018caj} in which the main results were presented. This paper will explain the results in more detail in addition to presenting other related results.

The paper is structured as follows. In Section~\ref{sec:differentcuts}, we will briefly review the amplituhedron and explain the geometry of the different cuts that we analyze in this paper. In Section~\ref{sec:deepestcuts}, we explore cuts which involve cutting $4L{-}4$ propagators. We derive expressions for these cuts and verify their correctness against known results. In Section~\ref{sec:deepcuts}, we derive the results for $2L{-}4$ cut propagators which in \cite{Arkani-Hamed:2018caj} were named the ``deepest cuts'' of the amplituhedron. Finally, in Section~\ref{sec:nontrivial}, we present a few preliminary results which involve solving nontrivial mutual positivity conditions. We consider the nontrivial deformations away from the deepest cuts, as well as generalized ladder cuts which are $n$-point extensions of the four-point results of \cite{IntoTheAmplituhedron}. 

\section{Geometry of the Amplituhedron}
\label{sec:differentcuts}
Although it was initially defined in terms of a generalization of the positive Grassmannian \cite{positiveGrass}, the amplituhedron can be defined entirely in terms of sign flip conditions on  intrinsically four-dimensional data \cite{Arkani-Hamed:2017vfh}. The external kinematic data for any massless scattering process is completely specified by the (null) external momenta $\{p_1,\ldots,p_n\}$ satisfying momentum conservation, and the helicities of the interacting particles. The external momenta can be completely specified by giving $n$ unconstrained momentum twistors $\left\lbrace Z_1, \ldots Z_n\right\rbrace$ as introduced in \cite{hodges}. In $\mathcal{N}=4$ sYM, it suffices to give the $\text{N}^k$MHV degree $k$ instead of specifying the individual helicities. Additionally, at $L$ loops the loop integration variables are given by $L$ lines $\mathcal{L}_\alpha=(AB)_\alpha$, $\alpha=1,\ldots,L$, each of which can be specified by two points say, $A_\alpha$ and $B_\alpha$. In terms of these variables, the amplituhedron is the region which satisfies the following conditions:
\bea
\label{eq:conditions}
&&\langle ii+1jj+1 \rangle >0,\qquad\forall i<j,\\
&&\nonumber \left\lbrace \langle 1234 \rangle, \langle 1235 \rangle, \dots \langle 123n\rangle \right\rbrace \text{has $k$ sign flips,}\\
&&\nonumber \langle (AB)_\alpha ii+1 \rangle >0 \qquad \forall \alpha \in \, \, \lbrace 1, \dots L\rbrace,\\
&&\nonumber{\{\langle (AB)_\alpha12\rangle,\ldots,\langle (AB)_\alpha 1n\rangle\}\qquad\text{has $k+2$ sign flips,}}\\
&&\nonumber \langle (AB)_\alpha (AB)_\beta \rangle > 0, \qquad \forall \,\, \alpha \,<\, \beta\,\,\text{and } \alpha, \, \beta \in \lbrace 1, \dots L\rbrace.
\eea
The $L$-loop integrand for the $\text{N}^k$MHV helicity configuration is the unique degree $4(k+L)$ differential form in $(Z_i,(AB)_\alpha)$ with logarithmic singularities on all boundaries of the space. For the MHV ($k=0$) helicity configuration, the sign flip conditions on the sequence $\{\langle (AB)_\alpha 1i\rangle\}_{i=2,\ldots,n}$ can be reformulated in a slightly different form in terms of the planes $\bar{i}\equiv(i{-}1ii{+}1)$ dual to the points $Z_i$ \cite{Arkani-Hamed:2017vfh}. For the MHV $L$-loop integrand we can equivalently impose the following set of conditions:
\bea
&&\langle ii+1jj+1 \rangle >0,\\
&&\nonumber \langle (AB)_\alpha \,\bar{i}\,\bar{j}\rangle>0,\qquad \forall \, i<j,\\
&&\nonumber \langle (AB)_\alpha (AB)_\beta \rangle > 0, \qquad \forall \,\, \alpha \,<\, \beta\,\,\text{and } \alpha, \, \beta \in \lbrace 1, \dots L\rbrace,
\eea
where we introduced the shorthand notation $\langle (AB)_\alpha\bar{i}\bar{j}\rangle\equiv\langle (AB)_\alpha (i{-}1ii{+}1)\cap(j{-}1jj{+}1)\rangle$ to denote the intersection of the planes $\bar{i}$ and $\bar{j}$. From these definitions, it is clear that solving the problem at $L$-loops amounts to solving the problem at one-loop together with the mutual positivity conditions $\langle (AB)_\alpha (AB)_\beta \rangle>0$.
\begin{figure}[t!]
\begin{center}
\includegraphics[scale=.5]{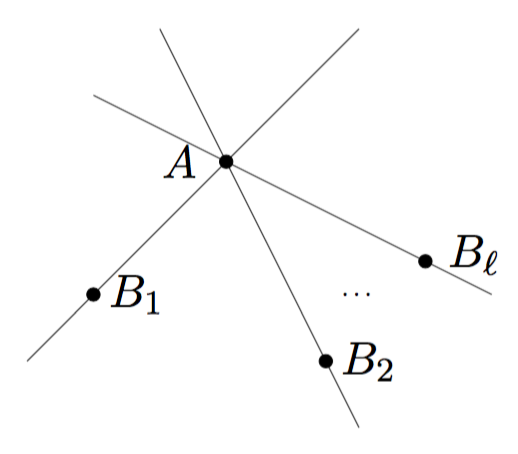}
\caption{Intersecting cut}
\label{fig:allInVertex}
\end{center}
\end{figure}
In this paper, we are interested in some faces of the amplituhedron which trivialize all mutual positivity constraints i.e., we approach the boundary where $\langle (AB)_\alpha(AB)_\beta\rangle=0$ for all $\alpha<\beta$. Generically, this set of constraints has two solutions which are related by parity i.e., the exchange of points$\leftrightarrow$planes. The first solution is a configuration of lines, all of which intersect at a single point $A$ as shown in Figure~\ref{fig:allInVertex}. We refer to this solution as the \emph{intersecting cut}.

\begin{figure}[t]
\begin{center}
    \includegraphics[scale=0.4]{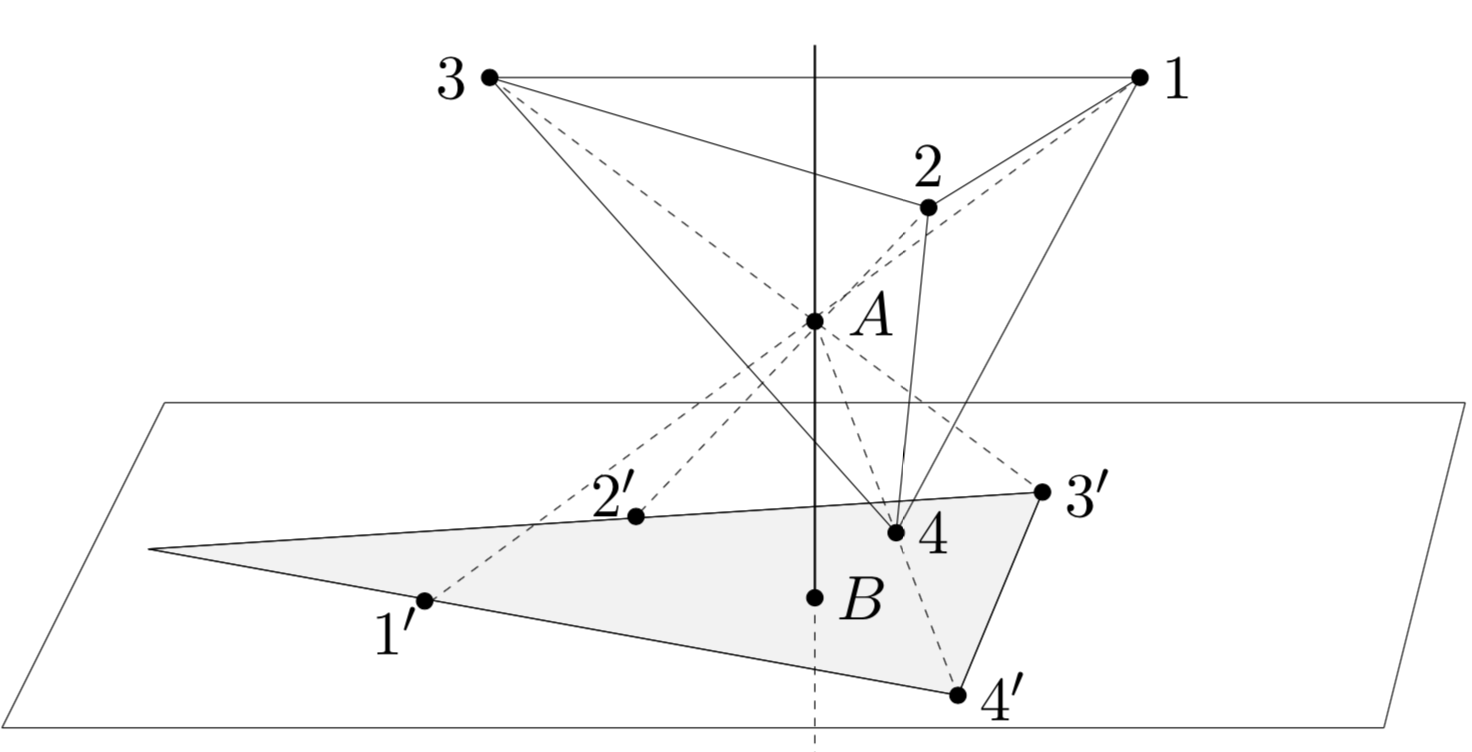}
\end{center}
\caption{Projection through $A$ at four points.}
\label{fig:Projection2}
\end{figure}

It is worthwhile to understand the counting of the number of degrees of freedom left on this boundary. We start with $L$ loops and hence $4L$ degrees of freedom. Making each loop pass through a point requires two constraints. Na\"{i}vely, this would require $2L$ constraints. However, the point at which all the lines intersect is not specified. Hence we only need $2L-3$ conditions, and the resulting form has degree $(2L+3)$. The remaining conditions on the loop lines are 
\bea
\label{eq:intersect}
\langle AB_\alpha \bar{i}\bar{j}\rangle >0.
\eea
These are completely independent of each other and the problem essentially reduces to $L$ copies of the one-loop problem. These inequalities determine the allowed locations of $A$ (which has three degrees of freedom) and also the allowed configuration of each line $AB_\alpha$ for a given $A$ (each $B_\alpha$ has two degrees of freedom left). We seek a cell decomposition of $A$-space such that for each cell in $A$ space, the geometry of $B_\alpha$ is fixed. By projecting through the common intersection point $A$ one possible one-loop configuration at, say, four points is given in Figure~\ref{fig:Projection2} (the full $L$-loop configuration is simply $L$ copies of this geometry).\footnote{Of course, at this point there is no reason to think that the configuration of Figure~\ref{fig:Projection2} is actually consistent with the inequalities defining the amplituhedron. However, as we shall demonstrate in Section~\ref{sec:deepcuts} this geometry does contribute to the intersecting cut.}
In this picture we see that $A$ lives inside a tetrahedron with vertices $Z_1,\ldots,Z_4$ while $B$ lives inside the triangle with vertices $Z_3',Z_4'$ and $(23)'{\cap}(14)'$. The triangulation of the intersecting cut is given by the set of all such configurations consistent with the inequalities defining the amplituhedron. Note that since the mutual positivity has been trivialized we expect that we will be able to write the canonical form such that it factorizes into a form for each cell in $A$ space and a product of forms for each loop $AB_\alpha$. Schematically, we have
\begin{equation}
\label{eq:factored}
\Omega^{\text{cut}}=\sum_A\Omega_A\prod_{\alpha=1}^L \Omega_{B_\alpha},
\end{equation}
where in this expression (and in many that follow) we suppress the measure of integration, which for the $L$-loop intersecting cut amounts to omitting the common factors $\langle A\mathrm{d}^3 A\rangle\prod_{\alpha=1}^L \langle AB_\alpha\mathrm{d}^2 B_\alpha\rangle$ from all expressions.

\begin{figure}[t]
\begin{center}
\includegraphics[scale=.6]{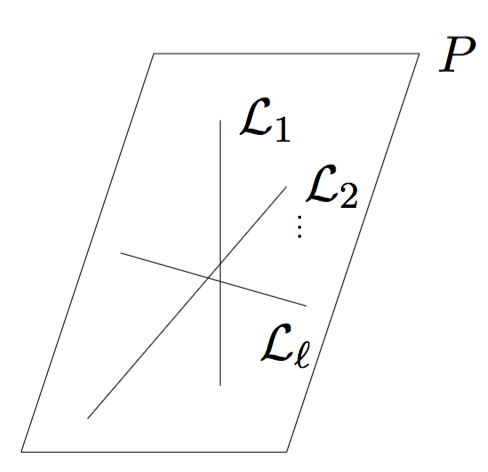}
\end{center}
\caption{All in plane cut}
\label{fig:allInPlane}
\end{figure}

The second solution to $\langle (AB)_\alpha (AB)_\beta \rangle = 0$ is the configuration in which all lines are coplanar but do not necessarily intersect at the same point shown in Figure~\ref{fig:allInPlane}. We refer to this solution as the coplanr cut.
\\
Let us denote the common plane by $(A_1A_2A_3)$. In this case, the remaining constraints read
\bea
\label{eq:planecondition}
\langle (A_1A_2A_3)B_\alpha \bar{i}\bar{j}\rangle > 0.
\eea
Since it is easier to work with points than to work with planes, we can dualize the above configuration. This involves the dual point $A^I =\epsilon^{IJKL}A_1^JA_2^KA_3^L$ . The dual of the condition in (\ref{eq:planecondition}) is 
\bea
\label{eq:plane}
\langle AB_\alpha ij\rangle >0.
\eea
We see that the dual configuration is now a set of lines $AB_\alpha$, all of which intersect at a point but satisfy $\langle AB_\alpha ij \rangle >0 $ rather than $\langle AB_\alpha \bar{i}\bar{j} \rangle > 0 $ as in (\ref{eq:intersect}). This demonstrates that the two cuts are distinct from each other.

To find the canonical form for the configuration in Fig.~\ref{fig:allInPlane}, we can find the canonical form associated to the dual inequalities (\ref{eq:plane}) and dualize the form, exchanging $Z\leftrightarrow W$. Here we are assuming that the dual of the canonical form of the dual region is equal to the canonical form of the original region. We refer the reader to \cite{Arkani-Hamed:2017tmz} for more details. Operationally, it is somewhat easier to compare our results for the coplanar cut to cuts of the corresponding parity conjugate, ``$\overline{\text{MHV}}$'' integrand, where by ``$\overline{\text{MHV}}$'' here we mean the integrand obtained by dualizing $Z\leftrightarrow W$. Note, however, that this is not quite the actual $\overline{\text{MHV}}$ integrand since this object is defined by setting $k=n{-}2$ in the full definition of the amplituhedron. The relationships can be summarized by
\begin{equation}
\begin{split}
&\text{MHV intersecting}\leftrightarrow``\overline{\text{MHV}}\text{''}\text{ coplanar }, \\
&``\overline{\text{MHV} }\text{''}\text{ intersecting}\leftrightarrow \text{MHV coplanar}.
\end{split}    
\end{equation}
Thus we can view the set of conditions $\langle AB_\alpha ij\rangle>0$ as defining the intersecting cut of the ``$\overline{\text{MHV}}$'' integrand, which is dual (by exhanging $Z\leftrightarrow W$) to the coplanar cut of the MHV integrand. Similarly, the MHV intersecting cut can be viewed as the dual of the ``$\overline{\text{MHV}}$'' coplanar cut. To keep notation consistent in the rest of this paper we will write all results in terms of the intersection point $A$, regardless of whether we are considering the intersecting or coplanar cut. Explicit formulae for the two coplanar cuts are obtained by dualizing expressions (\ref{eq:coplanarCut}) and (\ref{eq:intersectingCut}). Before solving these two cuts, however, we will first consider an even simpler set of geometries where the intersecting/coplanar lines satisfy additional constraints.\\

\section{$4L-4$ Cuts of Amplitudes}
\label{sec:deepestcuts}
\subsection{Intersecting cut}
\label{sec:cyclicpolytopecuts}
In this section, we will focus on a configuration of lines $(AB)_\alpha \,\, \alpha=1,\dots L$, all of which intersect at a common point $A$. Additionally, we will demand that some of them pass through the points $Z_i$. Let us suppose that $AB_\alpha$ for some $\alpha$ passes through $Z_1$. The constraints that this imposes are given by a special case of (\ref{eq:planecondition}), i.e. $\langle A1\bar{i}\overline{j}\rangle>0$. It is straightforward to show this implies that $\{\langle A123 \rangle,$ $\langle A134 \rangle$, $\dots,$ $\langle A1n2 \rangle \}$ must all have the same sign. Geometrically, this implies that after projecting through $Z_1$, the point $A$ lies in the polygon with vertices $\{\hat{Z}_2,\hat{Z}_3,\dots,\hat{Z}_n\}$ (where the hats indicate the projection through $Z_1$). We can thus express $A = c_2 Z_2+ \dots c_n Z_n$ with $c_i>0$. Similarly, for a line passing through $Z_{i_\alpha}B$ we have the constraint that $\langle Aii{+}1i{+}2 \rangle$, $ \langle Aii{+}2i{+}3 \rangle$, $\ldots$, $\langle Ain(-1) \rangle$, $\dots$ and  $\langle Ai(-(i{-}2))(-(i{-}1)) \rangle$ all have the same sign. In this case, we can write \begin{equation}A = -c_1 Z_1-c_2 Z_2 - \dots +c_{i+1}Z_{i+1}\dots+c_n Z_n, \end{equation} with $c_i>0$.

Thus each line $(AB)_\alpha$ which passes through some point $Z_{i_\alpha}$ imposes constraints on the possible positions of the intersection point $A$. These are all linear inequalities on the $\mathbf{P}^3$ in which $A$ lives. Therefore they cut out some polytope, provided the inequalities are mutually consistent. To check for the consistency, it suffices to keep track of the sign pattern in the expansion of $A$ in terms of the $Z_i$. For example, passing through $Z_1$ forces the pattern to be $(?++\dots +)$ or  $(?--\dots -)$ and $Z_2$ forces $(-?++\dots +)$ or $(+?--\dots -)$, where the $?$ means that there are no constraints on the sign of that coefficient. We will now demonstrate this in detail for a few examples.

Let us begin with the the simplest case of $n=4$ points and two loops. Here we have two lines $AB_1$ and $AB_2$, and we demand that these pass through $Z_1$ and $Z_2$. We can expand 
\bea
A = c_1 Z_1+c_2 Z_2 + c_3 Z_3 + c_4 Z_4.
\eea
Passing through $Z_1$ imposes the pattern $(?+++) \text{   or   } (?-\,-\,-\,)$ on the signs of the coefficients $c_i$,
\bea
c_2>0 \, ,c_3>0\, , c_4>0, \, \text{   or   } c_2<0 \, ,c_3<0\, , c_4<0. \,
\eea
Similarly, passing through $Z_2$ imposes the pattern $(-?++)$ or $(+?--)$.
We see that the only consistent patterns are $(-+++)$ or $(+---)$. These are equivalent up to an overall sign and we can write $A = -Z_1 + c_2 Z_2 + c_3 Z_3 + c_4 Z_4$. This is indeed a polytope as stated above. Namely, it is a tetrahedron with vertices $Z_2, Z_3, Z_4$ and $-Z_1$.

Still working with two loops, we can consider the configuration that results from demanding that the lines pass through $Z_1$ and $Z_3$. The patterns imposed on the $c_i$ are $(?+++) \text{ or } (?---)$ from $Z_1$ and $(--?+) \text{ or } (++?-)$ from $Z_3$. To obtain a consistent pattern from these, we would need to make one of the $c_i$ vanish. This results in a degenerate configuration and is not allowed for generic loop momenta. Thus there are no consistent patterns and the cut must vanish. We know that this is indeed the case, as shown in \cite{TheAmplituhedron}. We will also verify this and more general predictions in Section~\ref{sec:verification}. 

While still working at two loops, we can easily generalize the above results to arbitrary $n$. If the two lines pass through $Z_1$ and $Z_2$, then, we have
\bea
A  =-Z_1+c_2 Z_2 +c_3 Z_3 + \dots +c_n Z_n.
\eea
Hence $A$ is in the convex hull of $\lbrace Z_2, \dots, -Z_1\rbrace$ which we denote as $A\in$ Conv$\left[Z_2, Z_3,\dots, -Z_1 \right]$.

We can further generalize to the configuration of lines passing through $Z_i$ and $Z_j$ (with $i< j$), with the result that \begin{equation}A \in \,\mathrm{Conv}\left[Z_j, \dots Z_n, -Z_1, \dots -Z_{i-1}\right]\quad \text{and} \quad A \in \,\mathrm{Conv}\left[Z_i, Z_{i+1} \dots Z_{j-1}, Z_j\right],\end{equation} provided neither is degenerate. Finally, for the most general case in which $L$ lines $(AB)_1$, $\ldots$, $(AB)_L$ pass through $Z_{i_1}, \dots Z_{i_L}$, respectively,  the above discussion shows that we can have $A\in $ Conv $\left[Z_{i_L},Z_{i_L+1}, \dots, Z_n, -Z_1, \dots, -Z_{i_1-1}\right]$, $A\in $  Conv $\left[Z_{i_1}, \dots, Z_{i_2}\right]$, $A\in $ Conv $\left[Z_{i_2},\dots,Z_{i_3}\right]$, up to $A\in$  Conv $\left[Z_{i_{L-1}} \dots Z_{i_L}\right]$, barring degeneracy.

\subsection{Verification}
\label{sec:verification}
In this section, we will verify all predictions made in Section~\ref{sec:cyclicpolytopecuts} for two loops. We do this by computing the cuts directly from the two loop MHV integrand which can be expressed in terms of a cylic sum of the double pentagons introduced in \cite{Arkani-Hamed:2010gh}. We denote the following diagram as $(ijkl)$:
\begin{figure}[htb!]
\centering
\includegraphics[scale=.5]{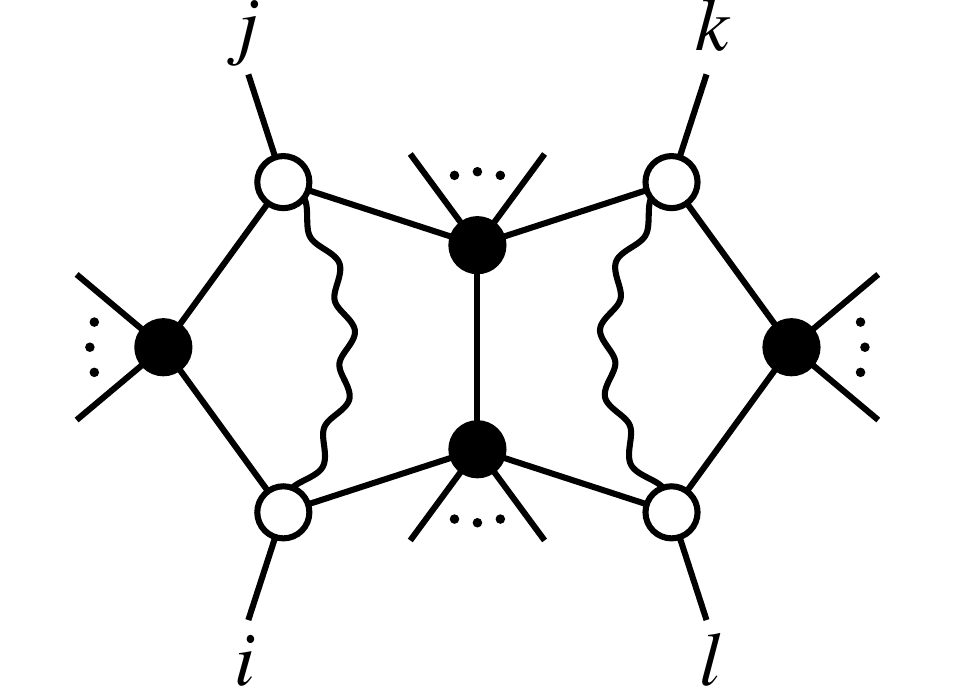}
\end{figure}

This picture represents the formula
\bea
(ijkl) = &&\frac{\langle AB \overline{i} \overline{j}\rangle}{\langle ABi{-}1i \rangle \langle ABii{+}1 \rangle \langle ABj{-}1j \rangle \langle ABjj{+}1 \rangle \langle ABCD \rangle}\\
&&\times\frac{\langle CD \overline{k}\overline{l}\rangle \langle ijkl \rangle }{\langle CDk{-}1k \rangle \langle
 CDkk{+}1\rangle \langle CDl{-}1l\rangle \langle CDll{+}1 \rangle},
\eea
where the two loop lines are $(AB)$ and $(CD)$. The MHV two-loop integrand can be expressed as a sum of double pentagons,
\bea
\label{eq:doublepentagon}
 \mathcal{A}_{n,\text{MHV}}^{\text{2-loop}} = &&\sum_{i<j<k<l<i} (ijkl).
\eea

We follow the same order as in the last section and begin with $n=4$. In this case the integrand can be expressed in terms of two double boxes 
\bea
\Omega_4 = \frac{\langle AB \mathrm{d}^2A \rangle \langle AB \mathrm{d}^2B \rangle \langle CD \mathrm{d}^2C \rangle\langle CD \mathrm{d}^2D \rangle \langle 1234 \rangle^3}{\langle AB14 \rangle\langle AB12 \rangle\langle AB34 \rangle \langle ABCD \rangle \langle CD12 \rangle \langle CD23 \rangle \langle CD34 \rangle}+\\
\frac{\langle AB \mathrm{d}^2A \rangle \langle AB \mathrm{d}^2B \rangle \langle CD \mathrm{d}^2C \rangle\langle CD \mathrm{d}^2D \rangle \langle 1234 \rangle^3}{\langle AB14 \rangle\langle AB12 \rangle\langle AB23 \rangle \langle ABCD \rangle \langle CD14 \rangle \langle CD23 \rangle \langle CD34\rangle}.
\eea
Taking the residue such that $AB$ passes through $Z_1$ and $CD$ through $Z_2$, we get\footnote{Henceforth where appropriate we will sometimes suppress the measure of loop integration.} 
\bea
\Omega_{4,\text{cut}} = \frac{\langle A\mathrm{d}^3A \rangle \langle 1234 \rangle^3}{\langle A123 \rangle \langle A134 \rangle \langle A412 \rangle \langle A423 \rangle},
\eea
where $A$ is the point of intersection of $AB$ and $CD$. This is precisely the canonical form for the simplex with vertices $Z_2, Z_3, Z_4, -Z_1$ as expected from Section~\ref{sec:cyclicpolytopecuts}.

We can also make $AB$ pass through $Z_1$ and $CD$ through $Z_3$. Taking residues appropriately, we find the residue on the cut vanishes
\bea
\Omega_{4,\text{cut}} = \frac{\langle 1234 \rangle^3}{\langle C142 \rangle\langle C134 \rangle \langle C312 \rangle \langle C234 \rangle} +\frac{ \langle 1234 \rangle^3}{\langle C142 \rangle\langle C123 \rangle \langle C314 \rangle \langle C234 \rangle} = 0,
\eea
exactly as predicted in Section~\ref{sec:cyclicpolytopecuts} and \cite{TheAmplituhedron}.

At five points, we next consider the cut where $AB$ passes through $Z_1$ and $CD$ through $Z_2$. Only three double pentagons contribute to this cut.  
\bea
&&(5123) \xrightarrow[D\rightarrow Z_2]{B\rightarrow Z_1} \frac{\langle C\mathrm{d}^3C \rangle\langle 5123 \rangle \langle 4512 \rangle \langle 1234 \rangle}{\langle C145 \rangle \langle C512 \rangle \langle C123 \rangle \langle C234 \rangle}\\
&&(5124) \xrightarrow[D\rightarrow Z_2]{B\rightarrow Z_1} -\frac{\langle C\mathrm{d}^3C \rangle\langle 5124 \rangle^2 \langle 2345 \rangle}{\langle C145 \rangle \langle C512 \rangle \langle C234 \rangle \langle C245 \rangle}\\
&&(4123)\xrightarrow[D\rightarrow Z_2]{B\rightarrow Z_1}\frac{\langle C\mathrm{d}^3C \rangle\langle 1234 \rangle^2 \langle 1345 \rangle}{\langle C134 \rangle \langle C145 \rangle \langle C123 \rangle \langle C234 \rangle}.\\
\eea
It is easy to check that this is indeed a triangulation  of the cyclic polytope with vertices $Z_2,Z_3,Z_4,Z_5,-Z_1$ as expected from Section~\ref{sec:cyclicpolytopecuts}.

More generally, at two loops if we have $AB$ passing through $Z_a$ and $CD$ passing through $Z_b$, the following double pentagons contribute:
\bea
&&(ajbl) \xrightarrow[D\rightarrow Z_b]{B\rightarrow Z_a}\frac{\langle A a \left( \overline{a}\cap \overline{j}\right)\rangle \langle A b \left( \overline{b}\cap \overline{l}\right) \langle abjl \rangle\rangle}{\langle A\overline{a}\rangle \langle Aaj-1j \rangle \langle Aajj+1 \rangle \langle A \overline{b}\rangle \langle Abl-1l\rangle \langle Abll+1\rangle}\\ 
&&(ajkb)  \xrightarrow[D\rightarrow Z_b]{B\rightarrow Z_a}\frac{\langle A a \left( \overline{a}\cap \overline{j}\right)\rangle \langle A b \left( \overline{k}\cap \overline{b}\right) \langle ajkb \rangle\rangle}{\langle A\overline{a}\rangle \langle Aaj-1j \rangle \langle Aajj+1 \rangle \langle A \overline{b}\rangle \langle Abk-1k\rangle \langle Abkk+1\rangle}\\
&&(iabl)  \xrightarrow[D\rightarrow Z_b]{B\rightarrow Z_a}\frac{\langle A a \left( \overline{i}\cap \overline{a}\right)\rangle \langle A b \left( \overline{b}\cap \overline{l}\right) \langle iabl \rangle\rangle}{\langle A\overline{a}\rangle \langle Aai-1i \rangle \langle Aaii+1 \rangle \langle A \overline{b}\rangle \langle Abl-1l\rangle \langle Abll+1\rangle}\\
&&(iakb)  \xrightarrow[D\rightarrow Z_b]{B\rightarrow Z_a}\frac{\langle A a \left( \overline{i}\cap \overline{a}\right)\rangle \langle A b \left( \overline{k}\cap \overline{b}\right) \langle iakb \rangle\rangle}{\langle A\overline{a}\rangle \langle Aai-1i \rangle \langle Aaii+1 \rangle \langle A \overline{b}\rangle \langle Abk-1k\rangle \langle Abkk+1\rangle}.
\eea
The form on this cut is then
\bea
\Omega = \sum_{j=a+1}^{b-1}\sum_{l=b+1}^{a-1} (ajbl)+ \sum_{j=a+1}^{b-2}\sum_{k=j+1}^{b-1} (ajkb)+ \sum_{l=b+1}^{a-2}\sum_{i=l+1}^{a-1} (iabl)+\sum_{i=b+1}^{a-2}\sum_{k=a+1}^{b-1} (iakb).
\eea
We expect this to be a triangulation corresponding to the sum of the forms for the two cyclic polytopes  Conv$[Z_a,\dots Z_b]$ and Conv$[Z_b,\dots Z_n, -Z_1, \dots, -Z_a]$. To verify this, we need the canonical form of a cyclic polytope. A triangulation of this form is given by $\Omega_1+\Omega_2$, where \cite{Arkani-Hamed:2017tmz}
\bea
\label{eq:cyclicpolyform}
\Omega_1 = \sum_{i=a+1}^{b-2} \left[aii+1b\right], \qquad\text{and}\qquad \Omega_2 = \sum_{i=b+1}^{a-2} \left[bii+1a\right],
\eea
where we define
\bea
\left[abcd \right] \equiv \frac{\langle abcd \rangle^3}{\langle Aabc\rangle \langle Abcd\rangle \langle Acda\rangle \langle Adab \rangle}.
\eea
We have verified up to $n=20$ that this prediction holds true in every case. However, the double pentagon expansion provides a triangulation of the two polytopes which is different from (\ref{eq:cyclicpolyform}). Furthermore, there is no obvious subset of terms in the double pentagon form which triangulates either the polytope Conv$[Z_a,\dots Z_b]$ or Conv$[Z_b,\dots Z_n, -Z_1, \dots, -Z_a]$ separately. This of course follows from the known fact that the double pentagon expansion of the integrand, although term-by-term local, is not a triangulation in the usual mathematical sense because it involves points living outside the amplituhedron. Understanding exactly how this representation of the integrand covers the amplituhedron, even on this special cut, is an interesting open question which we leave to future work. 
\subsection{Coplanar cut}
\label{sec:coplanardeepcut}
\begin{figure}[t!]
\begin{center}
\includegraphics[scale=.45]{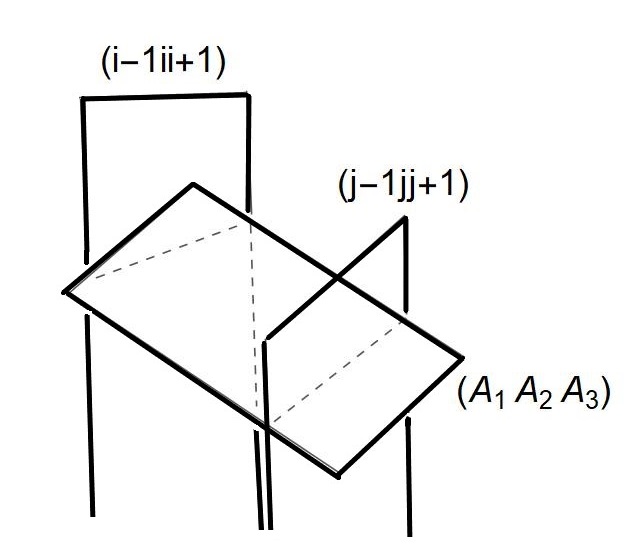}
\end{center}
\caption{The dotted lines are common plane $(A_1A_2A_3)$ intersecting $(i{-}1ii{+}1)$ and $(j{-}1jj{+}1)$.}
\label{fig:coplanardeep}
\end{figure}

In this section we will focus on the coplanar cut of the MHV integrand. Since we are considering coplanar lines, we cannot demand that they pass through the $Z_i$. This is impossible for generic configurations of external data. However, there exists a natural analog of making the lines $AB_\alpha$ pass through $Z_i$. Consider the planes $(i{-}1ii{+}1)$, which are dual to the points $Z_i$. These intersect the plane $(A_1A_2A_3)$ in lines as shown in Figure~\ref{fig:coplanardeep}.

We can identify the lines $(AB)_\alpha$ with these lines. To understand why this is a natural analog, it is helpful to look at the dual picture. Recall that the dual of a set of coplanar lines is a set of lines intersecting at a point. The dual of a line lying in the plane $(i{-}1ii{+}1)$ is a line passing through the point $Z_i$. Thus the dual of the configuration shown in Figure~\ref{fig:coplanardeep} is a set of lines intersecting at a point and passing through $Z_i, Z_j$ and $Z_k$. For the rest of this section, we will be working with the dual configuration and demanding the constraints $\langle ABij \rangle >0$ as explained in Section~\ref{sec:differentcuts}. We will denote the dual of the common plane $A_1A_2A_3$ by the point $A$. 

The coplanar cut is strikingly different from the intersecting cut. It lacks the rich structure of deeper cuts that we saw in Section~\ref{sec:cyclicpolytopecuts}. The first result which sets the two cuts apart is that we cannot make $AB_\alpha$ pass through non consecutive $Z_i$. To see this, it suffices to look at the constraints imposed by passing through $Z_a$ and $Z_b$ for $b>a$. Let us suppose that passing through $a$ imposes $\langle Aaij \rangle>0$. Passing through b then requires $\langle Abij \rangle <0$ since $a<b$ and we must have a consistent sign for $\langle Aabj \rangle$. Now, if there exists $c$ such that $a<c<b$ we have a contradiction, and therefore such a configuration of lines does not belong to the one-loop amplituhedron.

Consequently, configurations of lines passing through three or more of the $Z_i$ are also disallowed since this will necessarily involve two non consecutive $Z_i$. 
\section{$2L-4$ Cuts of Amplitudes}
\label{sec:deepcuts}
\begin{figure}[t]
\begin{center}
\begin{tabular}{|c|c|c|c|}
\hline $\ell$ & total $\#$ of topologies & possible contributions & $\%$ \\ \hline
4 & 8 & 4 & 50 \\ \hline
5 & 34 & 20 & 58.8 \\ \hline
6 & 229 & 146 & 63.8 \\ \hline
7 & 1873 & 1248 & 66.6 \\ \hline
8 & 19 949 & 13 664 & 68.5 \\ \hline
9 & 247 856 & 172 471 & 69.6 \\ \hline
10 & 3 586 145 & 2 530 903 & 70.6 \\ \hline
\end{tabular}
\end{center}
\caption{Number of topologies contributing on the cut through ten loops \cite{Arkani-Hamed:2018caj}.}
\label{fig:table1}
\end{figure}
\begin{figure}[t]
\includegraphics[scale=0.4]{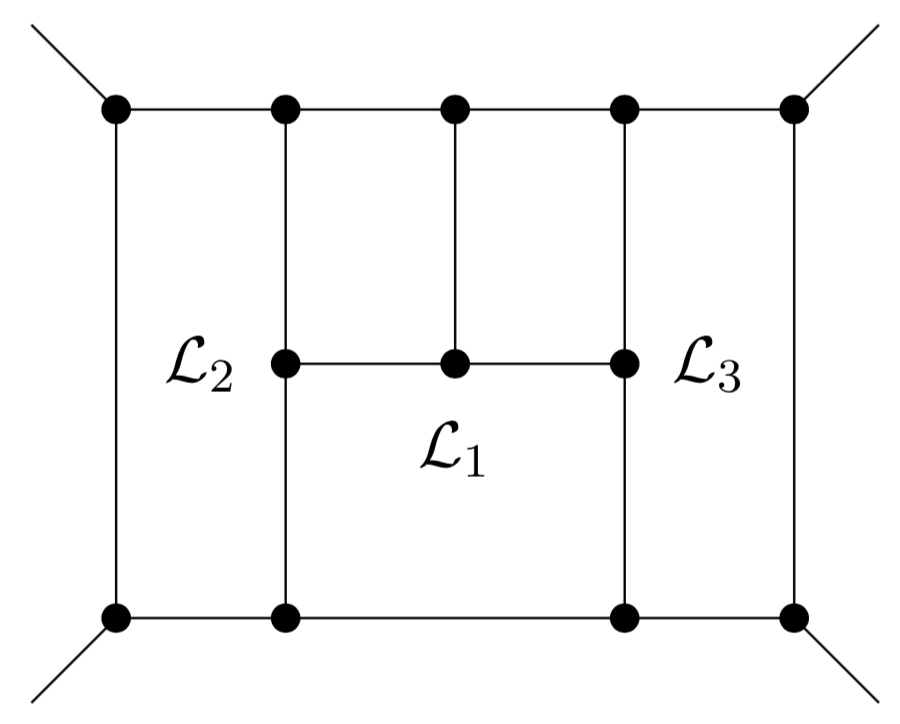}
\includegraphics[scale=0.35]{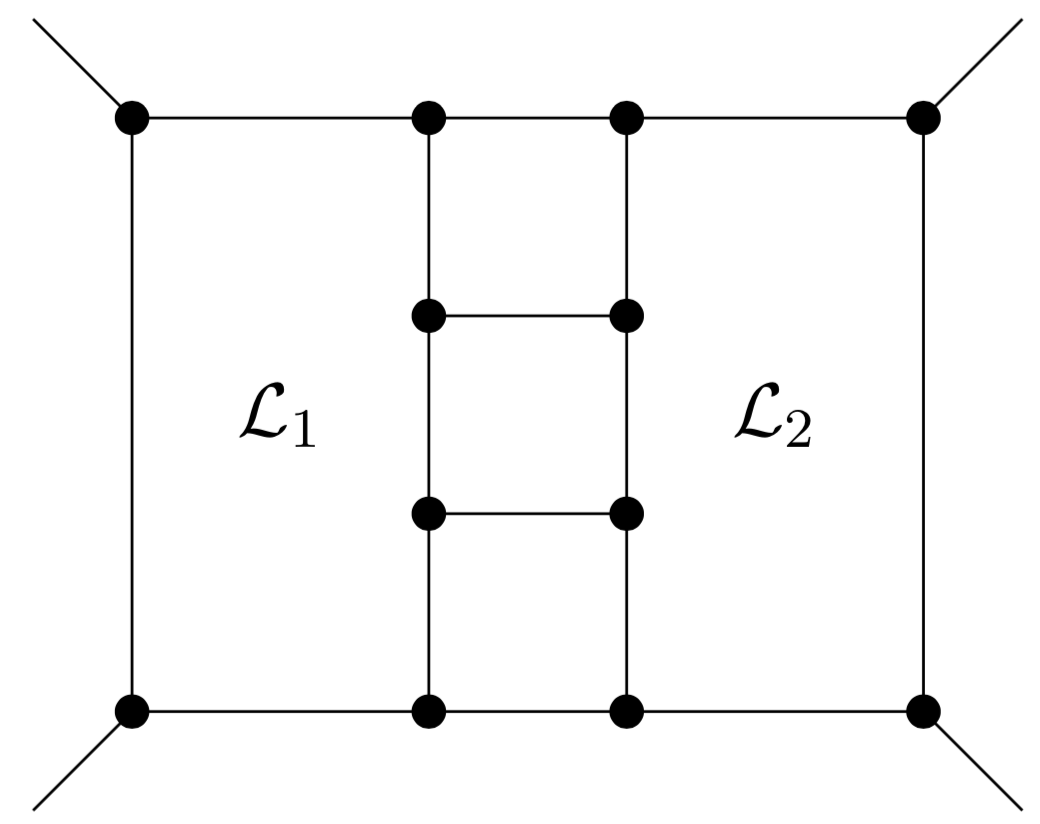}
\caption{The two local diagrams at five loops which have the necessary number of internal propagators but nevertheless do not contribute to the $(2L{-}4)$-dimensional cuts. Here we label the lines $\mathcal{L}_\alpha=(AB)_\alpha$.}
\label{fig:dci_diagrams}
\end{figure}

\noindent We now tackle the problem of finding the form for the cut $\langle (AB)_\alpha (AB)_\beta \rangle= 0$ with no other constraints imposed. As discussed in \cite{Arkani-Hamed:2018caj} this cut is hopelessly complicated from a local diagram expansion. We can be slightly more quantitative about the complexity of this cut by estimating how many local diagrams contribute at, say, $n=4$ points using known results available from the soft collinear bootstrap program \cite{Bourjaily:2011hi,Bourjaily:2015bpz,bourjaily2016}. From the ancillary files in \cite{bourjaily2016} the number of dual conformal invariant (DCI) integrals that have enough internal propagators to possibly contribute on the cut can be counted through ten loops, with the number of topologies given in Figure~\ref{fig:table1} which is taken from \cite{Arkani-Hamed:2018caj}. Note in particular that the \emph{total} number of diagrams is given by symmetrizing in all loop momenta and cycling through external labels. Of course, simply having enough internal propagators is not sufficient to say that a given diagram actually has support on our cuts, since there may be compensating DCI numerators which cancel some internal propagators and/or kill the residue. Thus, the numbers shown in the ``possible contributions'' column of Figure~\ref{fig:table1} are overestimates of the actual contributions, as can be seen by, for example, a more detailed consideration of the thirty-four topologies present at five loops: of these planar graphs twenty have at least the required seven internal propagators necessary to a priori contribute on the cut. However, of these twenty the two graphs shown in Figure~\ref{fig:dci_diagrams} have the associated DCI numerators 
\begin{equation}
N_1=-\langle1234\rangle^4\langle 12(AB)_1\rangle^2\langle(AB)_2(AB)_3\rangle^2,
\end{equation}
and
\begin{equation}
N_2=-\langle1234\rangle^4\langle (AB)_1(AB)_2\rangle^4,
\end{equation}
respectively. Therefore neither of these diagrams have nonzero residue on our cut, and the correct counting at five loops is eighteen rather than twenty.

\subsection{Four point problem}
We will first focus on the intersecting cut at four points. The inequalities for the line $AB_\alpha$ to be in the one-loop amplituhedron are $\langle AB_\alpha\overline{i}\overline{j}\rangle >0$. These reduce to
\begin{equation} 
\begin{split}
\label{eq:4ineq}
&\langle AB_\alpha12\rangle>0,\quad \langle AB_\alpha 13\rangle<0,\quad \langle AB_\alpha 14\rangle>0, \\ &\langle AB_\alpha23\rangle>0,\quad \langle AB_\alpha 24\rangle<0,\quad \langle AB_\alpha 34\rangle>0.
\end{split}
\end{equation}
The inequalities that result from the coplanar cut $\langle ABij\rangle >0$ are identical to (\ref{eq:4ineq}) except for the signs of $\langle AB_\alpha13\rangle$ and $\langle AB_\alpha24\rangle$. However, the form for the two inequalities is identical as the case of $n=4$ is too simple to distinguish between the two cuts. We can solve the system in (\ref{eq:4ineq}) explicitly by setting 
\begin{equation}
\label{eq:linear}
A=Z_1+a_2 Z_2+a_3 Z_3+a_4 Z_4,\quad B_\alpha=Z_1+x_\alpha Z_2+y_\alpha Z_3,
\end{equation} 
and solving the resulting inequalities for $a_2$, $a_3$, $a_4$,  $x_\alpha$ and $y_\alpha$. The resulting triangulation is the union of the following four regions: 
\begin{itemize}
\item $a_4 < 0 \qquad a_3 > 0 \qquad a_2 < 0 \qquad a_2 < x_\alpha < 0 \qquad 0 < y_\alpha < (a_3 x_\alpha)/a_2$,
\item $a_4>0 \qquad a_3<0 \qquad a_2<0 \qquad x_\alpha<a_2 \qquad y_\alpha>0 $,
\item $a_4>0 \qquad a_3>0 \qquad a_2>0 \qquad x_\alpha<0 \qquad 0<y_\alpha<a_3 x_\alpha/a_2 $,
\item $a_4>0 \qquad a_3>0 \qquad a_2<0 \qquad x_\alpha<a_2 \qquad y_\alpha>a_3 x_\alpha/a_2 $.
\end{itemize}
This determines the canonical form for the region of interest in terms of $a_2,a_3,a_4, x_\alpha\text{ and } y_\alpha$. It is trivial to take these expressions and rewrite them in terms of momentum twistors by solving the linear equations (\ref{eq:linear}) for all variables. We refer the reader to \cite{IntoTheAmplituhedron} for numerous example of writing down the canonical forms corresponding to regions defined by inequalities, and here give only the final expression for the four-point form:
\bea
\label{eq:fourPointResult}
\Omega_4^{(L)}=&&\frac{\langle1234\rangle^3}{\langle A123\rangle\langle A124\rangle\langle A134\rangle\langle A234\rangle}\times\\
&&\nonumber\left(\prod_{\alpha}\frac{\langle A123\rangle\langle A234\rangle}{\langle AB_\alpha 12\rangle\langle AB_\alpha 23\rangle\langle AB_\alpha34\rangle}+\prod_{\alpha}\frac{(-1)\langle A123\rangle\langle A124\rangle}{\langle AB_\alpha 12\rangle\langle AB_\alpha23\rangle\langle AB_\alpha14\rangle}\right.\\ 
&&\nonumber\left. +\prod_\alpha\frac{\langle A124\rangle\langle A134\rangle}{\langle AB_\alpha12\rangle\langle AB_\alpha34\rangle\langle AB_\alpha14\rangle}+\prod_\alpha\frac{(-1)\langle A134\rangle\langle A234\rangle}{\langle AB_\alpha23\rangle\langle AB_\alpha 34\rangle\langle AB_\alpha 14\rangle}\right),
\eea
where at four points there is only one form in $A$,
\begin{equation}\Omega^{(4)}_A=[1234]=\frac{\langle 1234\rangle^3}{\langle A123\rangle\cdots\langle A412\rangle},\end{equation}
which corresponds to the tetrahedron with faces $Z_i$ \cite{polytopes}. This clearly shows that the form in $A$, $\Omega_A$, is independent of the number of loops, $L$. 

\subsection{Five point coplanar cut}
At five points we can algebraically solve the inequalities $\langle AB_\alpha ij\rangle>0$ by parametrizing $A$ and $B_\alpha$ as above and triangulating the space of allowed common points $A$ for fixed geometries in $B_\alpha$. As the number of inequalities to solve becomes large for higher points, this approach becomes computationally intractable.
\begin{figure}[t!]
\begin{center}
\includegraphics[scale=0.35]{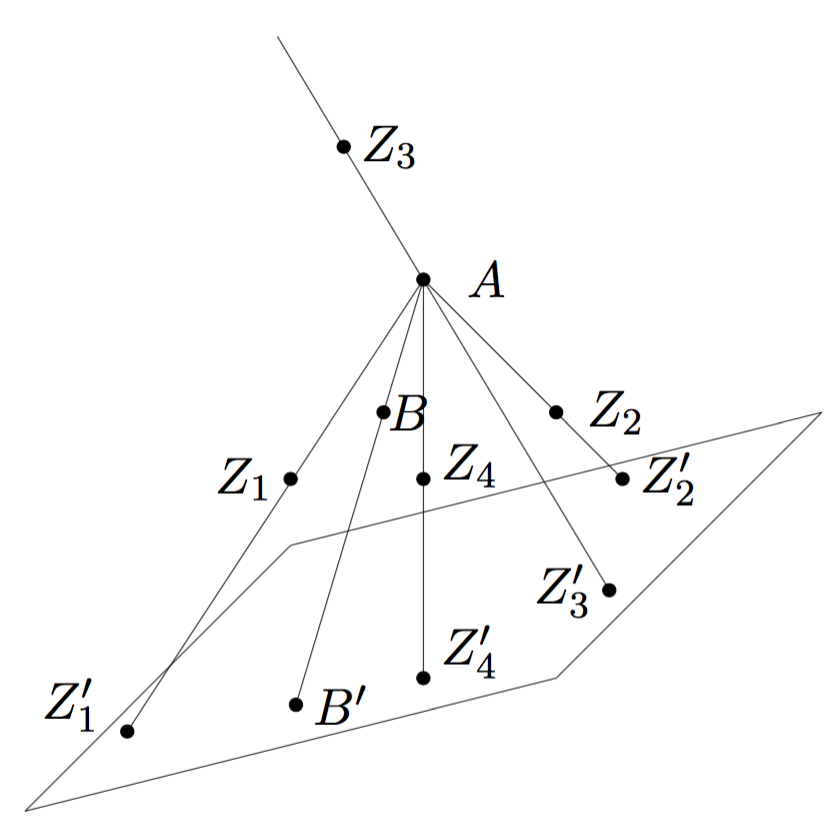}
\caption{Projecting through $A$ to get a two-dimensional configuration.}
\label{fig:projection}
\end{center}
\end{figure}
However, the geometry of the problem is quite simple: we have several intersecting lines with at most quadratic inequalities between them. This suggests that the pieces in the triangulation might in some sense be ``simple.'' To see if this is possible we seek an alternative procedure to solve the inequalities which is completely geometric rather than algebraic in nature. In fact, this reformulation of the problem is easy to find: to ``triangulate'' the space of allowed $AB_\alpha$ we should simply draw all configurations of points $\{Z_1, \ldots Z_n\}$ allowed by the inequalities $\langle AB_\alpha ij \rangle >0$. This is efficiently accomplished by first projecting the external data and the points $B_\alpha$ through the common intersection point $A$, whence we land on the two dimensional picture of Figure~\ref{fig:projection} where the bracket $\langle AB ij\rangle$ is positive if the point $B$ lies to the right of the line $(ij)$.

For a given configuration of projected positive external data $Z_1',\ldots,Z_n'$ (henceforth we omit the primes on projected variables) the conditions that $AB_\alpha$ is in the one-loop amplituhedron simply demand that the projected point $B_\alpha$ lies to the right of all lines $(ij)$, for $i<j$. This generates a list of allowed configurations in $A$ along with the corresponding regions in $B$ from which we can directly write down the forms.

\begin{figure}[t]
 \includegraphics[scale=0.5]{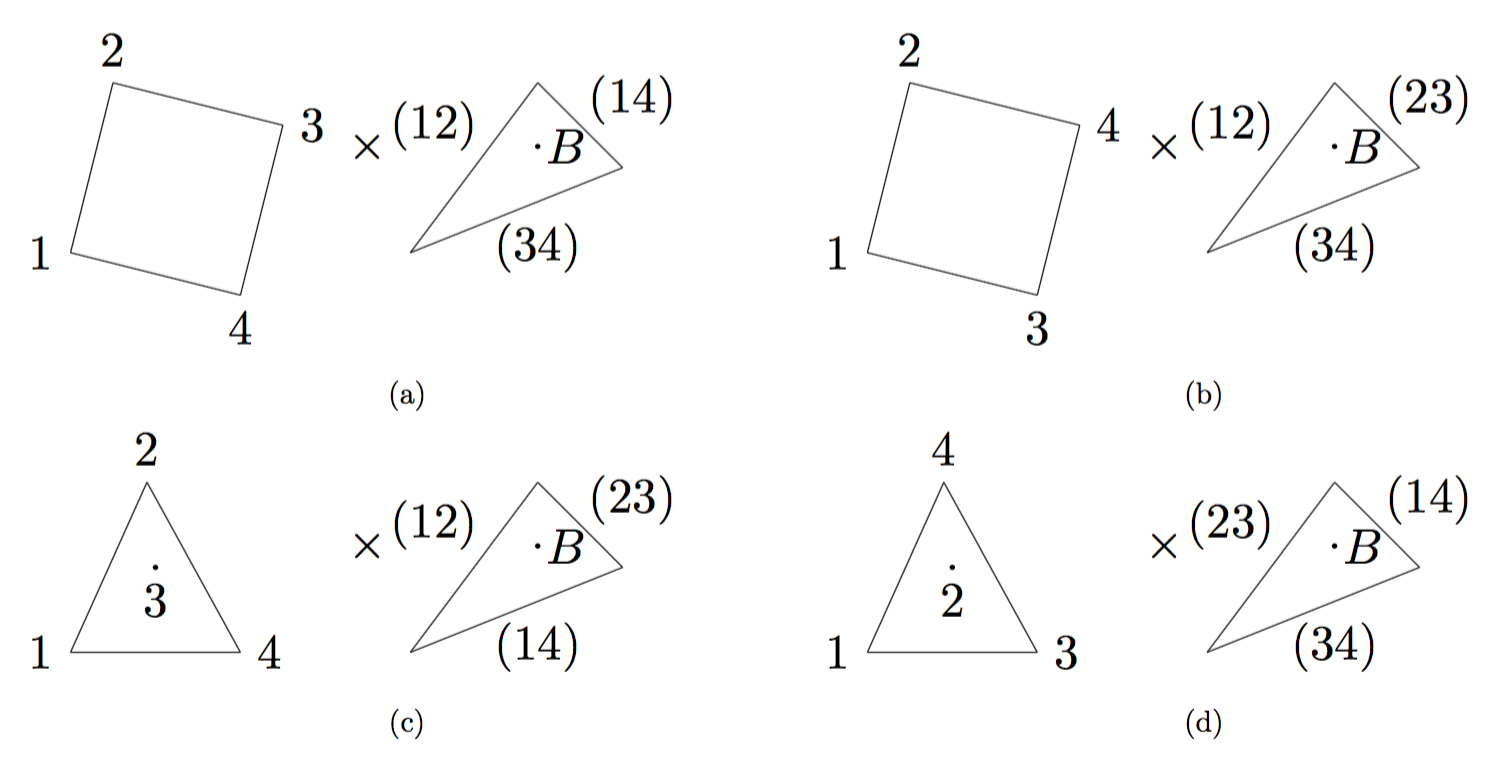}
 \caption{Four point configurations written as $(\text{configuration in $A$})$ and $(\text{allowed region in $B$})$.}
  \label{fig:fourPointConfigurations}
 \end{figure}

\begin{figure}[t]
\centering
\includegraphics[scale=0.55]{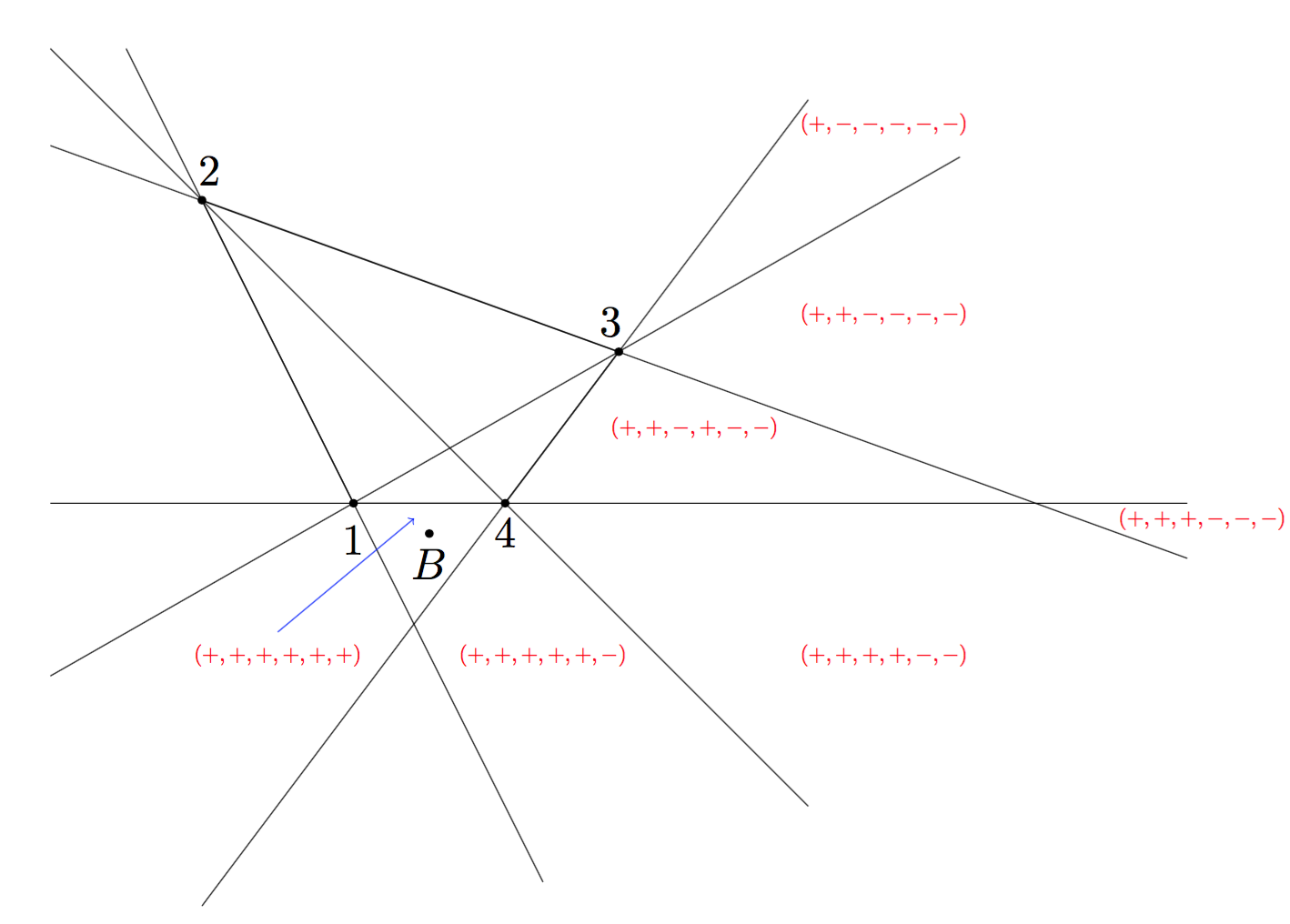}
\caption{Allowed regions in the projection plane for the point $Z_5'$, labelled by the sign sequence of (\ref{eq:sequence1}).}
\label{fig:fivePoint1}
\end{figure}

\begin{figure}[t]
\begin{subfigure}[b]{0.3\textwidth}
\includegraphics[scale=0.4]{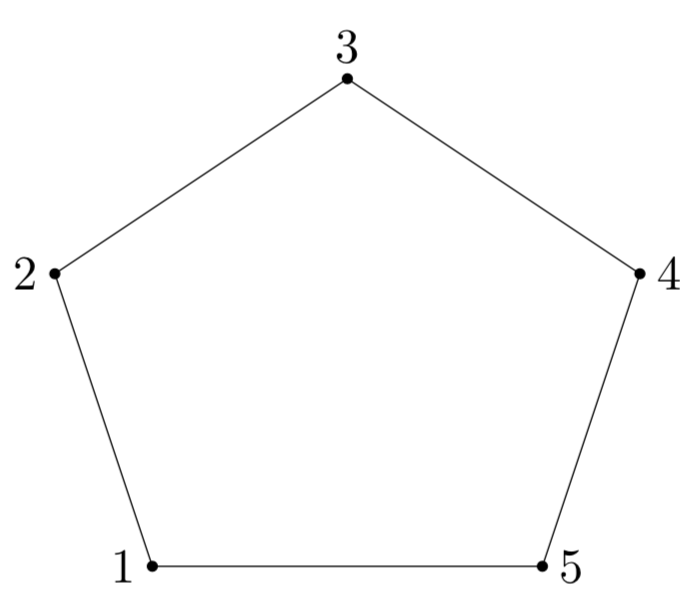}
\caption{}
\end{subfigure}
\begin{subfigure}[b]{0.3\textwidth}
\includegraphics[scale=0.68]{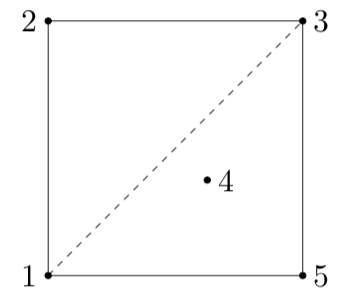}
\caption{}
\end{subfigure}
\begin{subfigure}[b]{0.3\textwidth}
\includegraphics[scale=0.43]{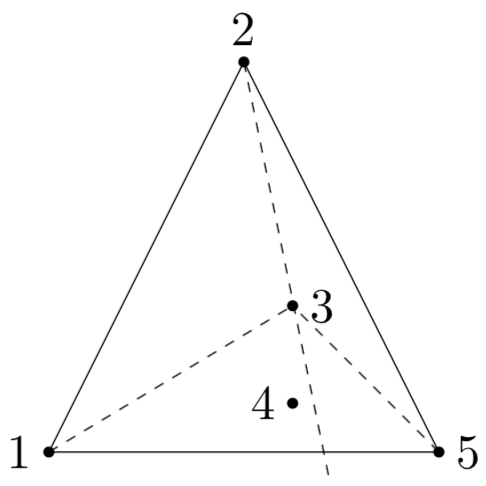}
\caption{}
\end{subfigure}
\caption{(a) Pentagonal configuration $12345$ with the region bounded by the lines $(12)(45)(15)$ for $B_\alpha$; (b) Quadrilateral configuration with the region bounded by lines (12)(15)(34) for $B_\alpha$; (c) Triangular configuration giving the same region for $B_\alpha$ as (b).}
\label{fig:pentagon}
\end{figure}

There are eight quadrilateral and eight triangular configurations for the four point case. Checking all possibilities against the inequalities $\langle AB_\alpha ij\rangle>0$ for $i<j=1,\ldots4$, we find four allowed configurations, displayed in Figure~\ref{fig:fourPointConfigurations} as the list of configurations in $A$ and the corresponding regions in $B$ where the inequalities are satisfied. From these pictures it is trivial to write down the corresponding canonical form, and we find term-by-term agreement with the algebraic approach of the previous section.
To solve the MHV coplanar (although here we are thinking of it as the ``$\overline{\text{MHV}}$'' intersecting) cut at five points, we can proceed by taking the four point configurations just obtained and adding a fifth point everywhere consistent with the additional five point inequalities $\langle AB i5\rangle>0$, for $i=1,\ldots,4$. For example, for configuration (a) of Figure~\ref{fig:fourPointConfigurations}, the point $Z_5$ can be added in any of the regions shown in Figure~\ref{fig:fivePoint1},
where in this picture we have labelled regions of the plane by the corresponding sign patterns of the sequence \begin{equation}\label{eq:sequence1}\{\langle A125\rangle,\langle A135\rangle,\langle A145\rangle,\langle A235\rangle,\langle A245\rangle,\langle A345\rangle\},\end{equation} and only configurations which give a nonzero allowed region for $B$ have been labelled. Although a priori this gives seven distinct configurations in $A$, in fact several of the configurations give identical allowed regions in $B$ and hence ``glue together'' naturally. If we complete this exercise for each four-point picture in Figure~\ref{fig:fourPointConfigurations}, the resulting list of configurations in $A$ and allowed regions for $B$ can be translated into the corresponding forms just as in the four point case. However, it is a less trivial exercise to compute the forms in $A$ corresponding to configurations of the $Z_i$. For example, one of the allowed configurations is the simple (projected) pentagon of Figure~\ref{fig:pentagon}(a), which gives for the point $B_\alpha$ the triangle bounded by lines $(12)(15)(45)$.
Here the codimension one boundaries in $A$ are obviously given by all deformations making three projected points collinear, so for the pentagon with ordered vertices $12345$ the poles of the form in $A$ are 
\begin{equation}
\{\langle A123\rangle,\langle A234\rangle,\langle A345\rangle,\langle A451\rangle,\langle A512\rangle\}.
\end{equation}
However, starting at five points we also find configurations such as in Figure~\ref{fig:pentagon}(b)-(c), both of which give the region for $B_\alpha$ bounded by the lines (12)(15)(34), where the pole structure of the form is not as obvious. The quadrilateral configuration of Figure~\ref{fig:pentagon}(b) has codimension one boundaries corresponding to the poles 
\[\{\langle A123\rangle,\langle A134\rangle,\langle A345\rangle,\langle A145\rangle,\langle A235\rangle,\langle A125\rangle\},\]
as can easily be seen by deforming the picture in all ways which make three points collinear. However, for the triangular configuration as drawn in Figure~\ref{fig:pentagon}(c) the relative orientation of points $Z_3$ and $Z_4$ inside the triangle is \emph{crucial} in reconstructing the form, and we must indicate whether the brackets $\{\langle A134\rangle,\langle A234\rangle,\langle A345\rangle\}$ are required to have definite signs in order to satisfy the inequalities. The codimension one boundaries of this cell correspond to those collinear limits which do not first flip any brackets which have definite sign. For Figure~\ref{fig:pentagon}(c) this gives, for example, the boundary structure corresponding to the poles
\[\{\langle A134\rangle,\langle A234\rangle,\langle A135\rangle,\langle A235\rangle,\langle A245\rangle\}.\]
The allowed regions in $B_\alpha$ can be classified by the pole structure of the associated form. In the four-point case we found all possible ``triangles'' with three poles in $B$ corresponding to the lines $(i{-}1i),(ii{+}1)$ and $(i{+}1i{+}2)$, and a priori at $n$-points one would anticipate finding triangles, quadrilaterals, etc. up to possibly $n$-gons for the allowed regions for $B_\alpha$. Indeed, adding a fifth point everywhere in the four-point configurations consistent with the additional five point inequalities yields both quadrilaterals and pentagons in $B_\alpha$. However, the corresponding sum of canonical forms for each cell does \emph{not} reproduce the correct integrand on this cut at any loop order. The reason for this discrepancy is simple: in addition to the inequalities $\langle AB_\alpha ij \rangle>0$ we must ensure that we are only keeping configurations that are consistent with having been projected from \emph{positive} data. To be more explicit, consider the following five-point configuration obtained from the procedure outlined above which is consistent with the inequalities $\langle AB_\alpha ij\rangle>0$ (here the point $Z_3$ must lie to the right of the line $(14)$ and to the left of the line $(25)$ to give the region in $B_\alpha$ shown)
\begin{equation}
\label{eq:illegal}
\includegraphics[scale=0.5]{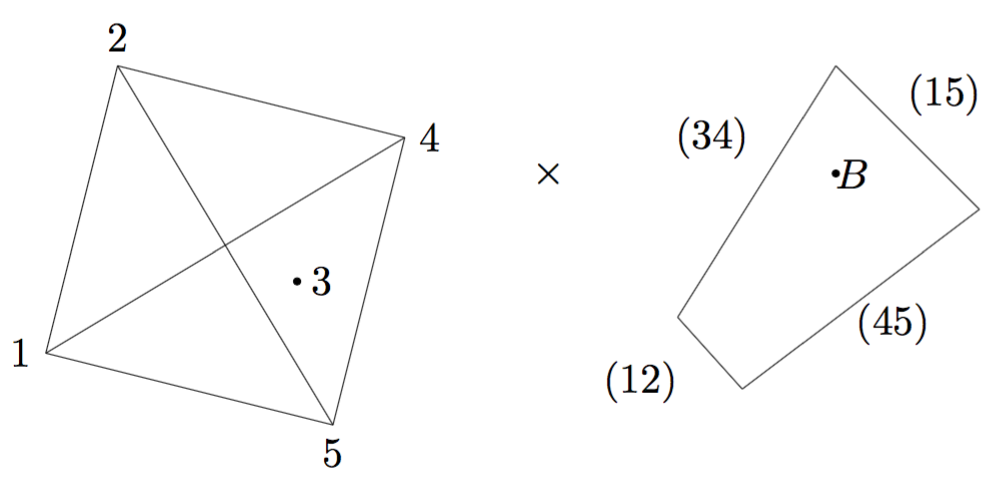}
\end{equation}
Na\"{i}vely this configuration contributes to the cut with a quadrilateral region for $B_\alpha$ with poles $\langle AB12\rangle,\langle AB34\rangle,\langle AB45\rangle$ and $\langle AB 15\rangle$. However, this configuration of projected $Z_i$ is actually \emph{inconsistent} with having been projected from \emph{positive} data, as a simple argument demonstrates. Namely, if we expand $Z_5$ in the basis $Z_1,\ldots,Z_4$ we have
\begin{equation} 
\label{eq:expansion}
Z_5=\alpha_4Z_4-\alpha_3Z_3+\alpha_2Z_2-\alpha_1Z_1,
\end{equation}
where the positivity of the variables $\alpha_i>0$ follows from the positivity of the external data. Expanding the bracket $\langle A135\rangle$ using (\ref{eq:expansion}) and noting that $\langle A123\rangle>0$ and $\langle A134\rangle<0$ (which are conditions defining this configuration) we see this implies this bracket is negative,
\begin{equation}
\langle A135\rangle=\alpha_4\langle A134\rangle-\alpha_3\langle A123\rangle<0,
\end{equation}
which is in contradiction to the configuration we have drawn, where $\langle A135\rangle>0$. Thus, the configuration (\ref{eq:illegal}) cannot be obtained by the projection of positive data. 

 If we cross-check the list of configurations obtained by adding a fifth point to the allowed four-point cases of Figure~\ref{fig:fourPointConfigurations} against the positivity constraints on the external data, the surprising result is the elimination of all geometries apart from triangles in $B_\alpha$. The complete set of configurations can be constructed out of the following list:
 \begingroup
\allowdisplaybreaks
\begin{align}
\label{eq:51}
&(12)(23)(34) & \raisebox{-35pt}{\includegraphics[scale=.35]{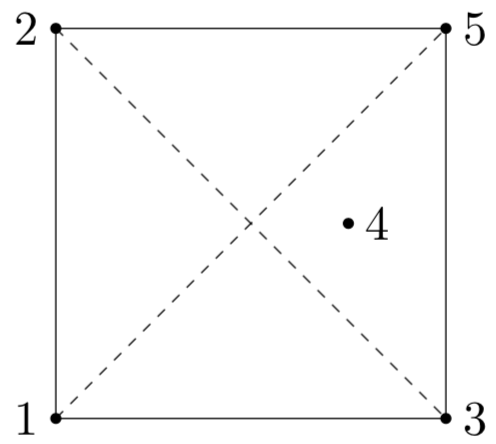}} &  \\
&(12)(23)(45) & \raisebox{-35pt}{\includegraphics[scale=.3]{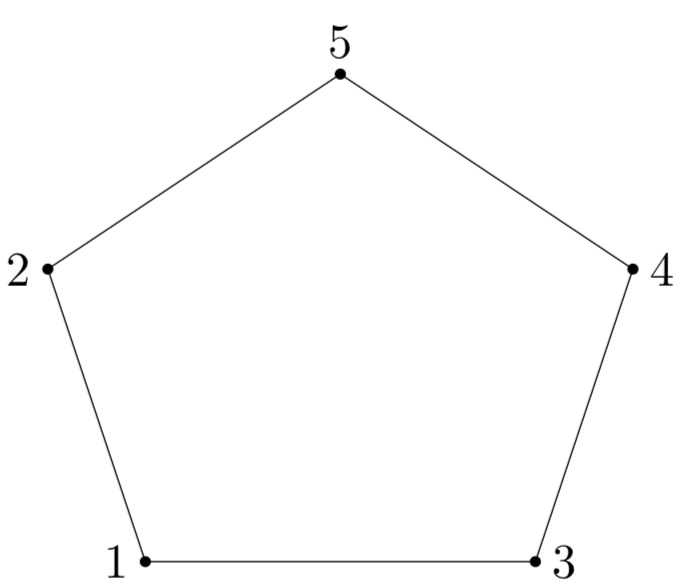}} &  \raisebox{-35pt}{\includegraphics[scale=.35]{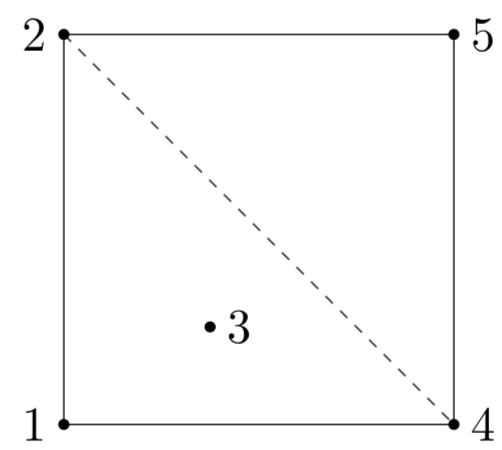}} \\
&(12)(23)(15) & \raisebox{-35pt}{\includegraphics[scale=.35]{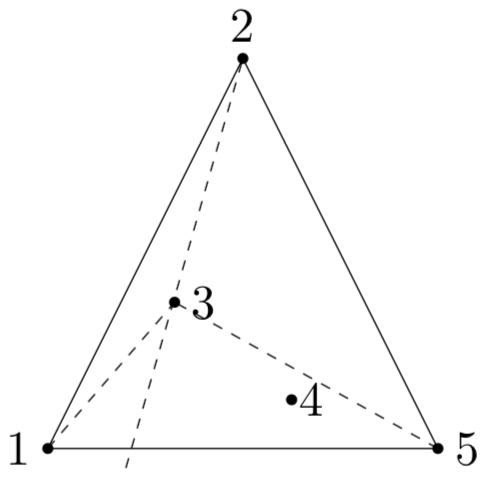}} &  \\
&(12)(34)(45) & \raisebox{-35pt}{\includegraphics[scale=.3]{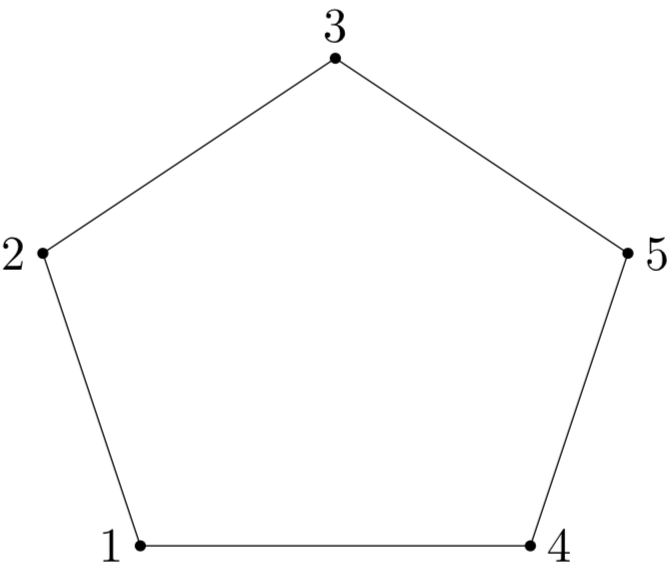}} & \raisebox{-35pt}{\includegraphics[scale=.35]{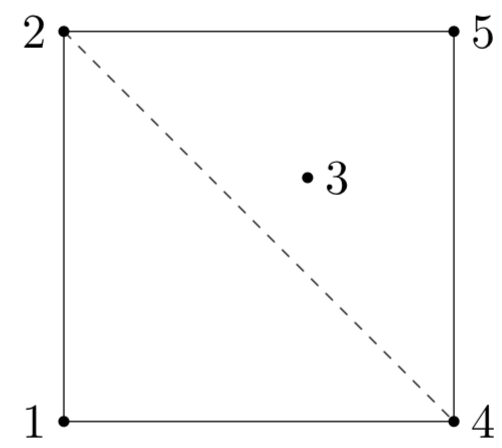}} \\
&(12)(34)(15) & \raisebox{-35pt}{\includegraphics[scale=.55]{quadrilateral2}} & \raisebox{-35pt}{\includegraphics[scale=.35]{triangle15}} \\
&(12)(45)(15) & \raisebox{-35pt}{\includegraphics[scale=.3]{pentagon3}} & \\
&(23)(34)(45) & \raisebox{-35pt}{\includegraphics[scale=.35]{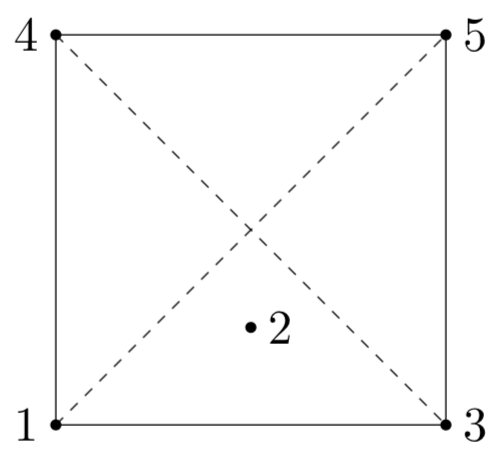}} & \\
&(23)(34)(15) & \raisebox{-35pt}{\includegraphics[scale=.35]{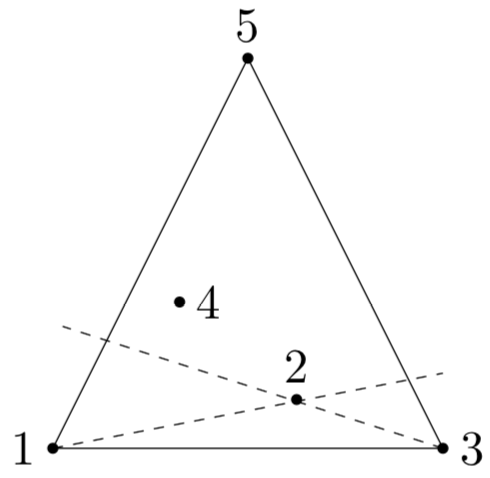}} &  (\text{no sign on $\langle A245\rangle$}) \\
&(23)(45)(15) & \raisebox{-35pt}{\includegraphics[scale=.35]{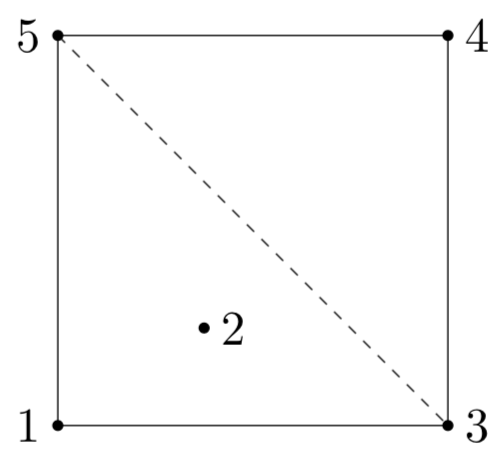}}  & \raisebox{-35pt}{\includegraphics[scale=.35]{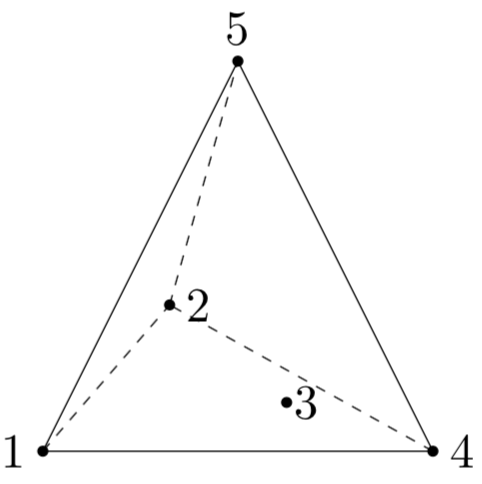}}\\
\label{eq:510}
&(34)(45)(15) & \raisebox{-35pt}{\includegraphics[scale=.35]{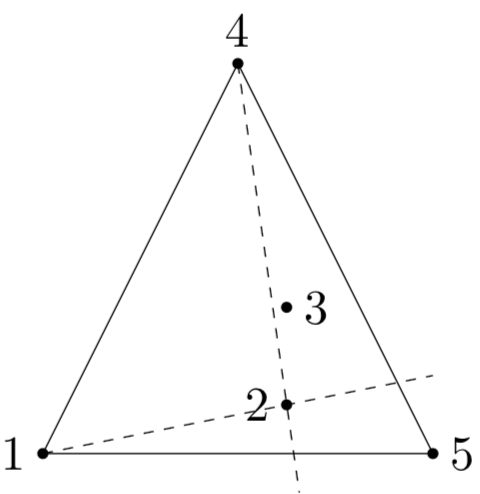}} & 
\end{align}
\endgroup
 In these results, we have indicated the regions in $B_\alpha$ satisfying the one-loop inequalities by the codimension one boundaries which are lines $(ii{+}1)$ in the projection through $A$. The full set of allowed configurations is given by adding all reflections across the line $(12)$ of the above list (disregarding duplicates), which is equivalent to requiring the consideration of both cases $\langle A123\rangle\gtrless0$. Alternatively, all possibilities can be generated by constructing, for each possible $B_\alpha$ region, all configurations consistent with the inequalities (with no requirements on any $\langle Aijk\rangle$ bracket); this leads to exactly the same set of allowed configurations as (\ref{eq:51})-(\ref{eq:510}), plus reflections across $(12)$. 
 
 As already mentioned, the key aspect of the five point results (\ref{eq:51})-(\ref{eq:510}) is that only triangles in $B_\alpha$ are found, despite there being no immediately obvious reason why quadrilaterals and pentagons are forbidden. In fact, if one repeats the above brute-force procedure to construct the complete set of six point geometries, the same simple result is found: only triangles in $B_\alpha$ satisfy the inequalities and are consistent with the positivity of external data. Although a deep explanation of why the positivity constraints demand triangle geometry for the $B_\alpha$ is at this point missing, in Appendix~\ref{sec:constraints} we discuss the precise nature of the constraints imposed on the projected data in slightly more detail. However, even without a satisfying explanation for this simplicity, we can immediately make an obvious ansatz: namely, for an \emph{arbitrary} number of particles, the geometry in $B_\alpha$ is still no more complicated than triangles! As we will see below this powerful hypothesis, checked by brute force at five and six points, allows use to solve the problem completely for any $n$ by a simple unitarity-inspired procedure. Since we know the allowed regions for the $B_\alpha$, we can obtain the corresponding $\Omega_A$ from the known two loop MHV integrands (\ref{eq:doublepentagon}) by taking residues. At higher points our triangle-hypothesis has been verified by matching our prediction for the cut against known expressions for the full integrand. 
\subsection{Coplanar cut for arbitrary multiplicities}
It was shown in Section~\ref{sec:coplanardeepcut} that the coplanar cut allowed only a limited number of deeper cuts. In particular, we cannot have any $\Omega_B$ which allows passing through more than two $Z_i$. We allow for all possible $\Omega_B$ with three factors of $\langle ABii+1 \rangle$ in the denominator and determine the corresponding $\Omega_A$. Surprisingly, this turns out to be the exact form on the cut for arbitrary $n$ and $L$. There can be three kinds of $\Omega_B$ with the following factors in the denominator.
\begin{itemize}
\item $\langle ABaa+1\rangle \,\langle ABbb+1\rangle \,\langle ABcc+1\rangle$
\item $\langle ABa-1a\rangle \,\langle ABaa+1\rangle \,\langle ABbb+1\rangle$
\item $\langle ABa-1a\rangle \,\langle ABaa+1\rangle \,\langle ABa+1a+2\rangle$
\end{itemize}
We can determine $\Omega_A$ for each of them by localizing the two-loop MHV integrand (\ref{eq:doublepentagon}) appropriately and computing the residues. Since the form for $A$ is independent of the number of loops, this gives us the form in $A$ for any number of loops.\\

\noindent\underline{Case 1: $aa{+}1 \text{-} bb{+}1 \text{-} cc{+}1$}\\
We can assume $a<b<c$ and no degeneracies (i.e $b\neq a+1, c\neq b+1, a\neq c+1$) and focus on the cut $\langle ABaa{+}1 \rangle = \langle ABbb{+}1 \rangle =0, \,\,\langle CDbb{+}1 \rangle = \langle CDcc{+}1 \rangle = 0$. The four double pentagons which contribute to this cut are $(a b b{+}1 c)$, $(a{+}1 b b{+}1 c)$, $(abb{+}1c{+}1)$, and $(a{+}1bb{+}1c{+}1)$. Their residues on this cut are
\bea
\label{eq:pentagonresidues1}
&&(abb+1c) \xrightarrow[B\rightarrow (Aaa{+}1 \cap Abb{+}1)]{D\rightarrow (Acc{+}1 \cap Abb{+}1)} \frac{\langle \overline{a}\,\overline{b}\,\overline{b{+}1}\,\overline{c} \rangle}{\langle A\overline{a}\rangle\langle A\overline{b}\rangle\langle A\overline{b{+}1}\rangle\langle A\overline{c}\rangle}\\
&&\nonumber(a+1bb+1c) \xrightarrow[B\rightarrow (Aaa{+}1 \cap Abb{+}1)]{D\rightarrow (Acc{+}1 \cap Abb{+}1)}  \frac{\langle \overline{a+1}\,\overline{b}\,\overline{b{+}1}\,\overline{c} \rangle}{\langle A\overline{a{+}1}\rangle\langle A\overline{b}\rangle\langle A\overline{b{+}1}\rangle\langle A\overline{c}\rangle}\\
&&\nonumber(abb{+}1c{+}1)\xrightarrow[B\rightarrow (Aaa{+}1 \cap Abb{+}1)]{D\rightarrow (Acc{+}1 \cap Abb{+}1)} \frac{\langle \overline{a}\,\overline{b}\,\overline{b{+}1}\,\overline{c{+}1} \rangle}{\langle A\overline{a}\rangle\langle A\overline{b}\rangle\langle A\overline{b{+}1}\rangle\langle A\overline{c{+}1}\rangle}\\
&&\nonumber(a+1bb+1c+1)\xrightarrow[B\rightarrow (Aaa{+}1 \cap Abb{+}1)]{D\rightarrow (Acc{+}1 \cap Abb{+}1)}  \frac{\langle \overline{a{+}1}\,\overline{b}\,\overline{b{+}1}\,\overline{c{+}1} \rangle}{\langle A\overline{a{+}1}\rangle\langle A\overline{b}\rangle\langle A\overline{b{+}1}\rangle\langle A\overline{c{+}1}\rangle}.
\eea
Here the bar represents the dual ($\overline{a} = (a{-}1aa{+}1)$). The sum of these four terms can be compactly written as
\bea
\label{eq:Octahedron}
\Omega_1 = \frac{\langle (Aaa{+}1 \cap Abb{+}1)cc{+}1 \rangle \langle a{-}1aa{+}1a{+}2 \rangle \langle b{-}1bb{+}1b{+}2\rangle \langle c{-}1cc{+}1c{+}2\rangle}{\langle A\overline{a}\rangle\langle A\overline{a{+}1}\rangle\langle A\overline{b}\rangle\langle A\overline{b{+}1}\rangle\langle A\overline{c}\rangle\langle A\overline{c{+}1}\rangle}
\eea
This is an octahedron with vertices \[\begin{split}&(\overline{a},\, \overline{b}, \, \overline{c}), (\overline{a},\, \overline{b}, \, \overline{c{+}1}), (\overline{a},\, \overline{b{+}1}, \, \overline{c}), (\overline{a},\, \overline{b{+}1}, \, \overline{c{+}1}), (\overline{a{+}1},\, \overline{b}, \, \overline{c}),\\ &(\overline{a{+}1},\, \overline{b}, \, \overline{c{+}1}), (\overline{a{+}1},\, \overline{b{+}1}, \, \overline{c}), \text{ and } (\overline{a{+}1},\, \overline{b{+}1}, \, \overline{c{+}1}).\end{split}\] The numerator puts a zero on all the other co-dimension 2 singularities. The facets are obvious from the expression. \\

\noindent\underline{Case 2: $a{-}1a \text{-} aa{+}1 \text{-} bb{+}1$}\\
A similar calculation shows that the form can be written as 
\bea
\label{eq:dualpolytope}
\Omega_2 = \frac{\langle Aabb{+}1\rangle \langle b{-}1bb{+}1b{+}2 \rangle \langle a{-}2a{-}1aa{+}1 \rangle \langle a{-}1aa{+}1a{+}2 \rangle}{\langle A\overline{a{-}1}\rangle \langle A\overline{a}\rangle \langle A\overline{a{+}1}\rangle \langle A\overline{b}\rangle \langle A\overline{b{+}1}\rangle}
\eea
This is a polytope with vertices \[\begin{split}&(\overline{a-1},\, \overline{a}, \, \overline{b}), (\overline{a-1},\, \overline{a}, \, \overline{b+1}), (\overline{a-1},\, \overline{a+1}, \, \overline{b}), (\overline{a-1},\, \overline{a+1}, \, \overline{b+1}),(\overline{a},\, \overline{a+1}, \, \overline{b}),\\ & \text{ and } (\overline{a},\, \overline{a+1}, \, \overline{b+1}) .\end{split}\] Again, the numerator puts a zero on all other co-dimension two singularities and the facets are obvious.\\

\noindent\underline{Case 3: $a{-}1a\text{-}aa{+}1\text{-}a{+}1a{+}2$}\\
Finally, we have
\bea
\label{eq:tetrahedron}
 \Omega_3 = \frac{\langle a{-}2a{-}1aa{+}1 \rangle \langle a{-}1aa{+}1a{+}2 \rangle \langle aa{+}1a{+}2a{+}3 \rangle}{\langle A\overline{a{-}1}\rangle \langle A\overline{a}\rangle \langle A\overline{a{+}1}\rangle \langle A\overline{a{+}2}\rangle}.
\eea
This is a tetrahedron with vertices $(\overline{a-1},\, \overline{a}, \, \overline{a+1}), (\overline{a-1},\, \overline{a}, \, \overline{a+2}),$ $(\overline{a-1},\, \overline{a+1}, \, \overline{a+2})$, and  $(\overline{a+1},\, \overline{a}, \, \overline{a+2})$.

 The full form at $L$-loops and arbitrary number of particles $n$ is given by summing over all possible triangles in $B_\alpha$. Note that the key aspect of this calculation was the fact that only triangles in $B_\alpha$ appear in the expansion (\ref{eq:factored}). If quadrilaterals and higher polygons appeared it would not, in general, be possible to fully fix the forms in $A$ just from the two-loop integrand. However, in this problem once we know the result on the cut can be expressed as a sum of triangles in $B_\alpha$ it is trivial to obtain the coefficients of the individual triangles. In particular, the triangles labelled by boundaries $(i{-}1i),(ii{+}1),(i{+}1i{+}2)$ are fixed by setting some set of the $B_\alpha=Z_i$ and the rest to $B_{\beta}=Z_{i+1}$. On this further cut of the integrand only this triangle can contribute. For example at two loops for the triangle $(12),(23),(34)$ the form in $A$ is fully fixed by solving the geometry when we cut $\langle AB_1 12\rangle=\langle AB_123\rangle=0$ and $\langle AB_223\rangle=\langle AB_2 34\rangle=0$ i.e., $B_1=Z_2$ and $B_2=Z_3$. For the triangles $(i{-}1i),(ii{+}1),(jj{+}1)$ we fix the coefficients by setting some $B_\alpha=Z_i$ and the rest to $B_\beta=(ii{+}1)\cap(Ajj{+}1)$. At two loops we can explicitly check that matching on this cut is sufficient to fix the coefficient of the triangle, matching on the other possible cuts $B_\alpha=Z_i$,$B_\beta=(i{-}1i)\cap(Ajj{+}1)$ and $B_\alpha=(i{-}1i)\cap(Ajj{+}1)$, $B_\beta=(ii{+}1)\cap(Ajj{+}1)$ is automatic. 

\begin{figure}[t]
\centering
\includegraphics[scale=0.6]{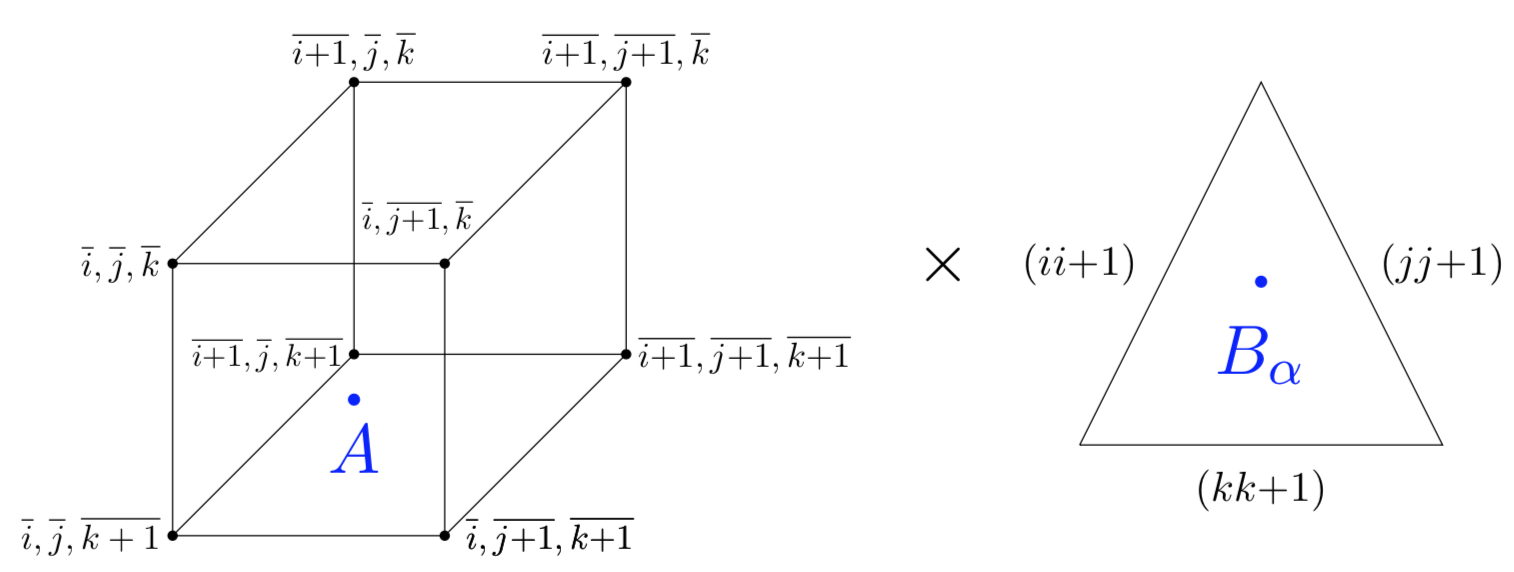}
\caption{The geometry of (the geometric dual of) the coplanar cut at $n$ points.}
\label{fig:dualForm}
\end{figure}

Proceeding in this way we obtain the full result for the $n$-point cut:
\bea
\label{eq:coplanarCut}
\nonumber &&\Omega_n^{(L)} = \frac{1}{L!}\sum_{i<j<k}\Bigg(\frac{\langle A(ii{+}1)\cap(Ajj{+}1)kk{+}1\rangle\langle i{-}1ii{+}1i{+}2\rangle\langle j{-}1jj{+}1j{+}2\rangle\langle k{-}1kk{+}1k{+}2\rangle}{\langle Ai{-}1ii{+}1\rangle\langle Aii{+}1i{+}2\rangle\langle Aj{-}1jj{+}1\rangle\langle Ajj{+}1j{+}2\rangle\langle Ak{-}1kk{+}1\rangle\langle Akk{+}1k{+}2\rangle}\\
&&\times \prod_{\alpha=1}^L\frac{\langle A(ii{+}1)\cap(Ajj{+}1)kk{+}1\rangle}{\langle AB_\alpha ii{+}1\rangle\langle AB_\alpha jj{+}1\rangle\langle AB_\alpha kk{+}1\rangle}\Bigg).
\eea
As discussed in Section~\ref{sec:differentcuts}, the final result (\ref{eq:coplanarCut}) is the correct formula for the $\overline{\text{MHV}}$ intersecting cut. To obtain the form for the MHV coplanar cut, we have to dualize (\ref{eq:coplanarCut}). As discussed in \cite{Arkani-Hamed:2018caj} the dual formula can be written \begin{equation}
\label{eq:dual}
\tilde{\Omega}_n^{(L)}=\sum_{i=1}^{n-2}\sum_{j=i+1}^{n-1}\sum_{k=j+1}^n \{i,j,k\}\wedge\prod_{\alpha=1}^L \frac{\mathrm{d}\mu_{\mathcal{L}_\alpha}\langle\langle P(i,j,k)\rangle\rangle}{\langle (AB)_\alpha ii{+}1\rangle\langle (AB)_\alpha jj{+}1\rangle\langle (AB)_\alpha kk{+}1\rangle},    
\end{equation}
where $\mathrm{d}\mu_{\mathcal{L}_\alpha}$ is the measure of the line $\mathcal{L}_\alpha$ on the plane $P$, and we define
\begin{equation}
    \{i,j,k\}\equiv \frac{\mathrm{d}\mu_P\langle\langle P(i,j,k)\rangle\rangle}{\langle Pi\rangle\langle Pi{+}1\rangle\langle Pj\rangle\langle Pj{+}1\rangle\langle Pk\rangle\langle Pk{+}1\rangle},
\end{equation}
and 
\begin{equation}
\langle \langle P(i,j,k)\rangle\rangle\equiv \langle ii{+}1 P{\cap}(jj{+}1)P{\cap}(kk{+}1)\rangle, 
\end{equation}
where $\mathrm{d}\mu_P$ is the measure of the plane $P$. In terms of the point $A$ and the planes $\bar{Z}_i$, the result (\ref{eq:dual}) can be schematically interpreted as in Figure~\ref{fig:dualForm}, where $\{i,j,k\}$ is the canonical form associated to a cube with facets associated to the lines $(ii{+}1)$ $(jj{+}1)$ and $(kk{+}1)$ and the form in $B_\alpha$ corresponds to a triangle in the plane with (the projections of) these lines. Note that, for example, in the case when $j=i{+}1$ and $k=i{+}2$ the geometry (and corresponding form) in (the dual of) $P$ smoothly degenerates to a tetrahedron.
\subsection{Verification of $\Omega_n^{(L)}$}
We have verified that the expression for $\Omega_n^{(L)}$ matches the coplanar cut of the two-loop MHV integrand up to $n=20$. We also verified that $\Omega_n^{(L)}$ reproduces the cut of the three-loop MHV integrand given in \cite{Arkani-Hamed:2010gh} up to (and including) $n=7$.

\subsection{Intersecting cut}

\subsubsection{Five points}
\begin{figure}[t]
    \centering
    \includegraphics[scale=0.4]{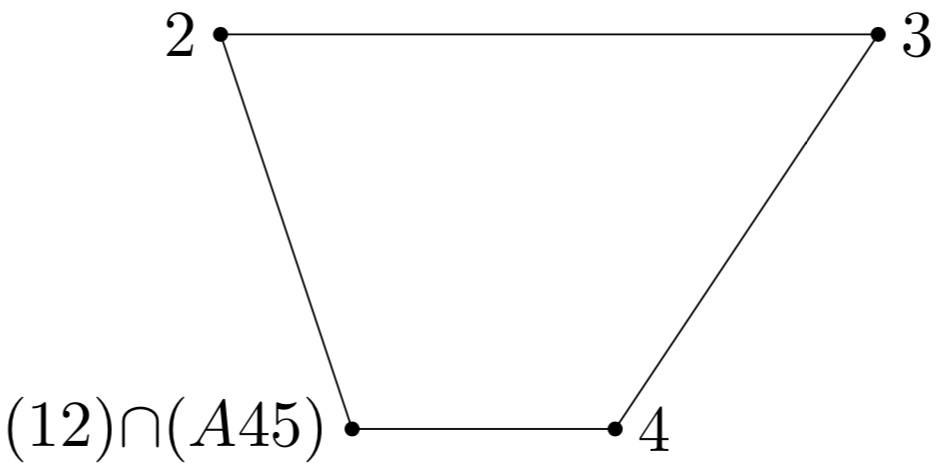}
    \caption{The region bounded by lines $(12),(23),(34)$ and $(45)$ whose canonical form is given by (\ref{eq:quadForm}).}
    \label{fig:Quad1}
\end{figure}

We now consider the MHV intersecting cut where all lines intersect in a common point $A$. Na\"{i}vely, one might hope that the simplicity of (\ref{eq:coplanarCut}) is mirrored in this cut as well. However, the lack of complexity in the coplanar cut arose from the fact that the only allowed regions in the $B_\alpha$ were triangles. This is clearly impossible for the intersecting cut due to the results of Section~\ref{sec:cyclicpolytopecuts} which show non-vanishing residues for the intersecting lines $(AB)_\alpha$ passing through any number of external points. It is also straightforward to verify that, for example, the three-loop five point integrand has a non-vanishing residue on the cut where $B_1=Z_2$ and $B_2=Z_4$ which no triangle in $B_\alpha$ can possibly reproduce. Instead, at five points we make the following ansatz:
\begin{equation}
\label{eq:ansatz}
\Omega=\sum_{\text{triangles }i} f_{t_i}(A)t_i(B_\alpha)+\sum_{\text{quadrilaterals }i}f_{q_i}(A)q_i,
\end{equation}
where the forms in $B_\alpha$ for the quadrilaterals have four poles and the numerators are determined by demanding unit leading singularities and vanishing on spurious singularities. For example, for the quadrilateral $q_1$ which corresponds to the region shown in Figure~\ref{fig:Quad1} bounded by the lines $(12),(23),(34),(45)$, the form is 
\begin{equation}
\label{eq:quadForm}
 q_1((12),(23),(34),(45))=\prod_\alpha\frac{\left(\langle AB_\alpha45\rangle\langle A123\rangle\langle A234\rangle-\langle AB_\alpha3(45){\cap}(A23)\rangle\right)}{\langle AB_\alpha12\rangle\langle AB_\alpha23\rangle\langle AB_\alpha34\rangle\langle AB_\alpha45\rangle}.
\end{equation}
This form gives the correct residues on $B_\alpha=Z_2,Z_3,Z_4$ and the numerator vanishes on spurious boundaries $B_\alpha=(12){\cap}(A34)$ and $B_\alpha=(23){\cap}(A45)$ (but does not vanish on $B_\alpha=(12){\cap}(A45)$). If we complete the exercise the forms for the additional quadrilaterals are given by: 
\begin{equation}
\begin{split}
q_2((12),(23),(34),(15))&=\prod_\alpha \frac{\langle AB_\alpha12\rangle\langle A135\rangle\langle A234\rangle+\langle AB_\alpha 34\rangle\langle A123\rangle\langle A125\rangle}{\langle AB_\alpha 12\rangle\langle AB_\alpha23\rangle\langle AB_\alpha34\rangle\langle AB_\alpha15\rangle},\\
q_3((12),(23),(45),(15))&=\prod_\alpha \frac{\langle AB_\alpha12\rangle\langle A145\rangle\langle A235\rangle+\langle AB_\alpha45\rangle\langle A123\rangle\langle A125\rangle}{\langle AB_\alpha12\rangle\langle AB_\alpha23\rangle\langle AB_\alpha 45\rangle\langle AB_\alpha 15\rangle}, \\
q_4((12),(34),(45),(15))&=\prod_\alpha \frac{\langle AB_\alpha15\rangle\langle A124\rangle\langle A345\rangle-\langle AB_\alpha34\rangle\langle A125\rangle\langle A145\rangle}{\langle AB_\alpha12\rangle\langle AB_\alpha 34\rangle\langle AB_\alpha45\rangle\langle AB_\alpha 15\rangle}, \\
q_5((23),(34),(45),(15))&=\prod_\alpha \frac{\langle AB_\alpha3(45){\cap}(A23)\rangle\langle A145\rangle-\langle AB_\alpha45\rangle\langle A135\rangle\langle A234\rangle}{\langle AB_\alpha23\rangle\langle AB_\alpha34\rangle\langle AB_\alpha45\rangle\langle AB_\alpha15\rangle}.
\end{split}
\end{equation} 
The coefficients of the quadrilaterals can be fixed from the two-loop result by considering particular cuts. For example, only the quadrilateral $q_1((12),(23),(34),(45))$ contributes on the cut where $B_1=Z_2$ and $B_2=Z_4$. This residue for the two-loop MHV integrand on the intersecting cut is 
\begin{equation}
\begin{split}
f_{q_1}(A)=&-\frac{\langle1234\rangle^2\langle1245\rangle}{\langle A123\rangle\langle A124\rangle\langle A145\rangle\langle A234\rangle}+\frac{\langle 1235\rangle\langle 1245\rangle\langle 1345\rangle}{\langle A123\rangle\langle A125\rangle\langle A145\rangle\langle A345\rangle}\\&-\frac{\langle 1234\rangle\langle 1235\rangle\langle 2345\rangle}{\langle A123\rangle\langle A125\rangle\langle A234\rangle\langle A345\rangle}
-\frac{\langle1234\rangle\langle1345\rangle\langle 2345\rangle}{\langle A123\rangle\langle A145\rangle\langle A234\rangle\langle A345\rangle}\\&-\frac{\langle 1245\rangle\langle2345\rangle^2}{\langle A125\rangle\langle A234\rangle\langle A245\rangle\langle A345\rangle}.
\end{split}
\end{equation}
However, this expression is deceptively complicated as a little algebra reveals that an equivalent form of the residue is simply
\begin{equation}
\label{eq:quadCoeff}
f_{q_1}(A)=\frac{\langle 1245\rangle^3}{\langle A124\rangle\langle A245\rangle\langle A451\rangle\langle A512\rangle}.
\end{equation}
An even faster way to fix (or alternatively double-check the derivation just given) the coefficient of $q_1((12),(23),(34),(45))$ is by considering the following cut of the three-loop five point integrand available in local form in~\cite{Arkani-Hamed:2010gh}: if we set $B_1=Z_2,B_2=Z_3,B_3=Z_4$ (which again isolates the coefficient of the quadrilateral) it is easily verified that the residue of the three-loop form on this cut is exactly (\ref{eq:quadCoeff}). The rest of the cuts are just as trivial; introducing the shorthand notation
\begin{equation}
[abcd]=\frac{\langle abcd\rangle^3}{\langle Aabc\rangle\cdots \langle Adab\rangle},
\end{equation}
the coefficients of the additional quadrilaterals are
\begin{equation}
\label{eq:quadCuts}
f_{q_2}(A)=[1345],\qquad f_{q_3}(A)=[2345], \qquad f_{q_4}(A)=[1234],\qquad f_{q_5}(A)=[1235].
\end{equation}
To fix the triangle coefficients we need only demand consistency on additional cuts. If we cut $B_1=Z_1$ and $B_2=Z_2$, the triangle with edges $(12),(23),(15)$ as well as the quadrilaterals $q_2$ and $q_3$ contribute. Therefore, we demand that the residue on the cut, which is
\begin{equation} 
\begin{split}
&\frac{\langle 1234\rangle^2\langle 1235\rangle}{\langle A123\rangle\langle A125\rangle\langle A134\rangle\langle A234\rangle}+\frac{\langle 1234\rangle\langle 1235\rangle\langle 1245\rangle}{\langle A123\rangle\langle A125\rangle\langle A145\rangle\langle A234\rangle}\\
&-\frac{\langle1234\rangle^2\langle1245\rangle\langle A135\rangle}{\langle A123\rangle\langle A125\rangle\langle A134\rangle\langle A145\rangle\langle A234\rangle}+\frac{\langle1245\rangle^2\langle 2345\rangle}{\langle A125\rangle\langle A145\rangle\langle A234\rangle\langle A245\rangle},
\end{split}
\end{equation}
matches the sum of the forms corresponding to the triangle $t_3$ and quadrilaterals $q_2,q_3$, 
\begin{equation} 
f_{t_3}(A)+f_{q_2}(A)+f_{q_3}(A).
\end{equation}
Using (\ref{eq:quadCuts}) this fixes the form in $A$ for the triangle $t_3$ to be surprisingly simple:
\begin{equation} 
f_{t_3}(A)=[1235].
\end{equation}
Checking all such cuts fixes the rest of the triangle coefficients. It is trivial to verify that at three loops the coefficients of all triangles and quadrilaterals are the same as at two loops. The final result at five points is:
\begingroup
\allowdisplaybreaks
\begin{align}
\label{eq:fivePoint}
\Omega_5=&+[1234]\prod_\alpha\frac{\langle A123\rangle\langle A234\rangle}{\langle AB_\alpha12\rangle\langle AB_\alpha23\rangle\langle AB_\alpha 34\rangle} \\
&+[5123]\prod_\alpha\frac{(-1)\langle A123\rangle\langle A125\rangle}{\langle AB_\alpha12\rangle\langle AB_\alpha23\rangle\langle AB_\alpha 15\rangle} \notag\\
&+[1245]\prod_\alpha\frac{\langle A125\rangle\langle A145\rangle}{\langle AB_\alpha12\rangle\langle AB_\alpha45\rangle\langle AB_\alpha 15\rangle} \notag\\
&+[2345]\prod_\alpha\frac{\langle A234\rangle\langle A345\rangle}{\langle AB_\alpha23\rangle\langle AB_\alpha34\rangle\langle AB_\alpha 45\rangle} \notag\\
&+[3451]\prod_\alpha\frac{(-1)\langle A345\rangle\langle A145\rangle}{\langle AB_\alpha34\rangle\langle AB_\alpha45\rangle\langle AB_\alpha 15\rangle} \notag\\
&+[1245]\prod_\alpha\frac{\langle AB_\alpha45\rangle\langle A123\rangle\langle A234\rangle-\langle AB_\alpha3(45){\cap}(A23)\rangle\langle A124\rangle}{\langle AB_\alpha12\rangle\langle AB_\alpha23\rangle\langle AB_\alpha34\rangle\langle AB_\alpha45\rangle} \notag\\
&+[3451]\prod_\alpha \frac{\langle AB_\alpha12\rangle\langle A135\rangle\langle A234\rangle+\langle AB_\alpha 34\rangle\langle A123\rangle\langle A125\rangle}{\langle AB_\alpha 12\rangle\langle AB_\alpha23\rangle\langle AB_\alpha34\rangle\langle AB_\alpha15\rangle} \notag\\
&+[2345]\prod_\alpha \frac{\langle AB_\alpha12\rangle\langle A145\rangle\langle A235\rangle+\langle AB_\alpha45\rangle\langle A123\rangle\langle A125\rangle}{\langle AB_\alpha12\rangle\langle AB_\alpha23\rangle\langle AB_\alpha 45\rangle\langle AB_\alpha 15\rangle} \notag\\
&+[1234]\prod_\alpha \frac{\langle AB_\alpha15\rangle\langle A124\rangle\langle A345\rangle-\langle AB_\alpha34\rangle\langle A125\rangle\langle A145\rangle}{\langle AB_\alpha12\rangle\langle AB_\alpha 34\rangle\langle AB_\alpha45\rangle\langle AB_\alpha 15\rangle} \notag\\
&+[5123]\prod_\alpha \frac{\langle AB_\alpha3(45){\cap}(A23)\rangle\langle A145\rangle-\langle AB_\alpha45\rangle\langle A135\rangle\langle A234\rangle}{\langle AB_\alpha23\rangle\langle AB_\alpha34\rangle\langle AB_\alpha45\rangle\langle AB_\alpha15\rangle}.\notag
\end{align}
\endgroup
This has been directly checked against the two and three-loop integrands evaluated on the intersecting cut. Note that all triangles of the form $((i{-}1i),(ii{+}1),(i{+}1i{+}2))$ appear in this expression, while the five triangles not of this form do not contribute at five points.

\subsubsection{Six points}
At six points it can be verified that on the cut $B_1=Z_2,B_2=Z_4,B_3=Z_5$ the three-loop integrand has nonzero residue. This implies that at the very least pentagons are necessary, since in our factorized ansatz only the pentagon with edges $((12),(23),(34),(45),(56))$ can possibly contribute on this cut. Writing down the general ansatz 
\begin{equation}
\label{eq:sixPointAnsatz}
\Omega_6=\sum_{\text{triangles }i} f_{t_i}(A)t_i(B_\alpha)+\sum_{\text{quadrilaterals }i}f_{q_i}(A)q_i(B_\alpha)+\sum_{\text{pentagons } i} f_{p_i}(A)p_i(B_\alpha),
\end{equation}
it is clear that once the forms in $A$ multiplying the pentagons are fixed it will be trivial to determine the forms for the quadrilaterals and triangles simply by demanding consistency on lower dimensional cuts. For example, once we compute cuts of the three-loop integrand and find that the coefficients of $p_1(12,23,34,45,56)$ and $p_2(12,23,34,45,16)$ are given by
\begin{equation}
f_{p_1}(A)=[1256]\qquad\text{and}\qquad f_{p_2}(A)=[4561],
\end{equation}
we can look at the two-loop integrand and cut $B_1=Z_2$ and $B_2=Z_4$, where only these two pentagons and the quadrilateral $q_1(12,23,34,45)$ contribute: 
\begin{equation}
\text{residue on cut}=f_{q_1}(A)+f_{p_1}(A)+f_{p_2}(A),
\end{equation}
which implies $f_{q_1}(A)=[1245]$, which is exactly the coefficient of this quadrilateral at five points. From these results we can immediately guess (and subsequently verify) the pattern: for the quadrilaterals \[(i{-}1i,ii{+}1,i{+}1i{+}2,i{+}2i{+}3)\] the corresponding forms in $A$ are $[i{-}1ii{+}2i{+}3]$, for the pentagons $(i{-}1i,\ldots,i{+}3i{+}4)$ the forms are $[i{-}1ii{+}3i{+}4]$ and for triangles \[(i{-}1i,ii{+}1,i{+}1i{+}2)\] the forms are $[i{-}1ii{+}1i{+}2]$. Checking the set of these cuts fixes the coefficients of all pentagons at six points as well as all quadrilaterals except those \emph{not} of the form \[(i{-}1i,ii{+}1,i{+}1i{+}2,i{+}2i{+}3),\] e.g., the quadrilateral $(12,23,34,56)$. However, it is easy to verify that all such quadrilaterals of this type, as well as the triangles not of the form $(i{-}1i,ii{+}1,i{+}1i{+}2)$, do not contribute to the integrand. For example, consider the cut $B_1=Z_2,B_2=Z_3$ of the two-loop integrand. Na\"{i}vely the following geometries contribute:
\begin{equation}
\begin{split}
\text{residue on cut}=&t(12,23,34)+q(12,23,34,45)+q(12,23,34,56)+q(12,23,34,16)\\
&+p(12,23,34,45,56)+p(12,23,34,45,16)+p(12,23,34,56,16).
\end{split}
\end{equation}
However if we substitute the known forms in $A$ we find this kills the coefficient of $q(12,23,34,56)$
\begin{equation}
\begin{split}
&\text{residue on cut}=[1234]+[1245]+f_q(12,23,34,56)+[6134]+[1256]+[6145]+[5634]\\
&\implies f_q(12,23,34,56)=0.
\end{split}
\end{equation}
A similar argument kills the quadrilateral $q(12,23,45,56)$ and all quadrilaterals of this type. The final expression for the six point integrand at $L$ loops is:
\begin{equation}
\begin{split}
&\sum_{i=1}^6 [i{-}1ii{+}1i{+}2]\prod_{\alpha=1}^L \frac{\langle Ai{-}1ii{+}1\rangle\langle Aii{+}1i{+}2\rangle}{\langle AB_\alpha i{-}1i\rangle\langle AB_\alpha ii{+}1\rangle\langle AB_\alpha i{+}1i{+}2\rangle}\\
&+\sum_{i=1}^6 [i{-}1ii{+}2i{+}3]\prod_{\alpha=1}^L \frac{N_{\text{quadrilateral}}(i)}{\langle AB_\alpha i{-}1i\rangle\langle AB_\alpha ii{+}1\rangle\langle AB_\alpha i{+}1i{+}2\rangle\langle AB_\alpha i{+}2i{+}3\rangle}\\
&+\sum_{i=1}^6 [i{-}1ii{+}3i{+}4]\prod_{\alpha=1}^L \frac{N_{\text{pentagon}}(i)}{\langle AB_\alpha i{-}1i\rangle\langle AB_\alpha ii{+}1\rangle\langle AB_\alpha i{+}1i{+}2\rangle\langle AB_\alpha i{+}2i{+}3\rangle\langle AB_\alpha i{+}3i{+}4\rangle},
\end{split}
\end{equation}
where $N_{\text{quadrilateral}}(i)$ and $N_{\text{pentagon}}(i)$ are the (unique) numerators which have unit leading singularities on codimension two boundaries such as $B_\alpha=Z_i,Z_{i+1},Z_{i+2},Z_{i+3}$ and vanish on spurious singularities such as $B_\alpha=(i{-}1i){\cap}(Ai{+}1i{+}2)$. We give explicit expressions for the form for the $k$-gon below. 

\subsubsection{Arbitrary multiplicities}
\begin{figure}
\includegraphics[scale=0.5]{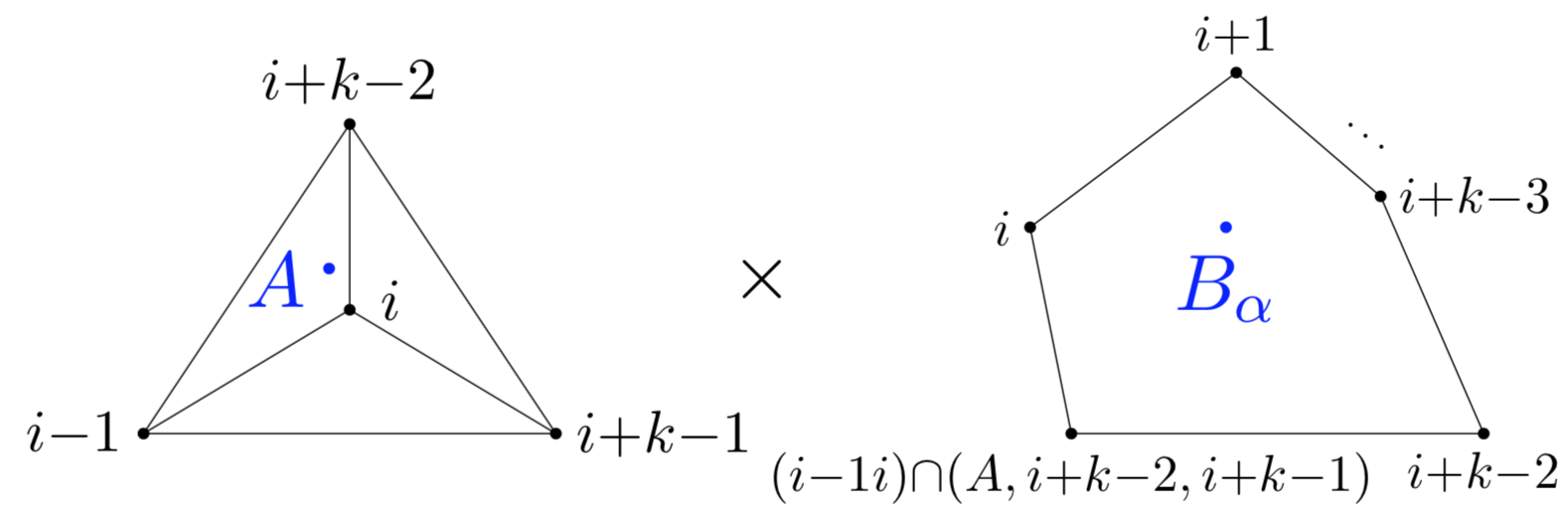}
\caption{$n$-point geometry for the intersecting cut.}
\label{fig:nPoint}
\end{figure}

From the six point result it is clear what our ansatz should be at $n$ points: all triangles, quadrilaterals, pentagons, $\ldots,$ up to $(n{-}1)$-gons which have only consecutive poles contribute on the cut. The form in $A$ for the $i$th $k$-gon is given by $[i{-}1,i,i{+}k{-}2,i{+}k{-}1]$ where $(i{-}1,i)$ labels the first edge and $(i{+}k{-}2,i{+}k{-}1)$ labels the last edge. The form is then 
\begin{equation}
\begin{split}
\Omega_n=&\sum_{i=1}^n [i{-}1ii{+}1i{+}2]\prod_{\alpha=1}^L \frac{\langle Ai{-}1ii{+}1\rangle\langle Aii{+}1i{+}2\rangle}{\langle AB_\alpha i{-}1i\rangle\langle AB_\alpha ii{+}1\rangle\langle AB_\alpha i{+}1i{+}2\rangle}\\
&+\sum_{i=1}^n [i{-}1ii{+}2i{+}3]\prod_{\alpha=1}^L \frac{N_{\text{quadrilateral}}(i)}{\langle AB_\alpha i{-}1i\rangle\langle AB_\alpha ii{+}1\rangle\langle AB_\alpha i{+}1i{+}2\rangle\langle AB_\alpha i{+}2i{+}3\rangle}\\
&+\sum_{i=1}^n [i{-}1ii{+}3i{+}4]\prod_{\alpha=1}^L \frac{N_{\text{pentagon}}(i)}{\langle AB_\alpha i{-}1i\rangle\langle AB_\alpha ii{+}1\rangle\langle AB_\alpha i{+}1i{+}2\rangle\langle AB_\alpha i{+}2i{+}3\rangle\langle AB_\alpha i{+}3i{+}4\rangle} \\
&+\cdots \\
&+\sum_{i=1}^n [i{-}1,i,i{+}n{-}3,i{+}n{-}2]\prod_{\alpha=1}^L \frac{N_{(n{-}1)-\text{gon}}(i)}{\langle AB_\alpha i{-}1i\rangle\langle AB_\alpha ii{+}1\rangle\cdots\langle AB_\alpha i{+}n{-}3,i{+}n{-}2\rangle},
\end{split}
\end{equation}
or more succinctly 
\begin{equation}
\label{eq:NpointMHV}
\Omega_n=\sum_{k=1}^{n-1}\sum_{i=1}^n [i{-}1,i,i{+}k{-}2,i{+}k{-}1]\prod_{\alpha=1}^L\frac{N_{k-\text{gon}}(i)}{\langle AB_\alpha i{-}1i\rangle\langle AB_\alpha ii{+}1\rangle\cdots\langle AB_\alpha i{+}k{-}2,i{+}k{-}1\rangle}.
\end{equation}
It is straightforward to verify that assuming (\ref{eq:NpointMHV}) is true at e.g., seven points is consistent with computing cuts of the two- and three-loop integrands, even without having the explicit form of the hexagons in $B_\alpha$. In fact, however, it is trivial to obtain the forms for any $k$-gon either using the procedure outlined in \cite{ArkaniHamed2015} or alternatively by simple triangulation. For a $k$-gon with the vertices \[Z_i,Z_{i+1},\ldots,Z_{i{+}k{-}2},(i{-}1i){\cap}(A,i{+}k{-}2,i{+}k{-}1)\] an expression for the form is given by
\begin{equation} 
\Omega_{k\text{-gon}}=\sum_{j=2}^{k-2}\frac{\langle Aij\widehat{j{+}1}\rangle^2}{\langle Aij\rangle\langle Aj\widehat{j{+}1}\rangle\langle A\widehat{j{+}1}i\rangle},\end{equation} 
where we define
\begin{equation}
\hat{Z}_{j+1}=\left\{\begin{array}{lr} Z_{j+1}, & j\neq k-2, \\
(i{-}1i){\cap}(Aijj{+}1), & j=k-2.\end{array}\right.
\end{equation}
The final expression for the intersecting cut is then:
\begin{equation}
\label{eq:intersectingCut}
\Omega_n=\sum_{k=1}^{n-1}\sum_{i=1}^n [i{-}1,i,i{+}k{-}2,i{+}k{-}1]\prod_{\alpha=1}^L\left(\sum_{j=2}^{k-2}\frac{\langle Aij\widehat{j{+}1}\rangle^2}{\langle Aij\rangle\langle Aj\widehat{j{+}1}\rangle\langle A\widehat{j{+}1}i\rangle}\right).
\end{equation}
Geometrically the solution can be described as $(\text{tetrahedron in }A)\times(\text{polygon in }B_\alpha)$ as in Figure~\ref{fig:nPoint}, which is directly reproduced from \cite{Arkani-Hamed:2018caj}.

\subsection{Verification of intersecting cut}
The result (\ref{eq:intersectingCut}) has been checked against the expressions for the two- and three-loop MHV integrands given in \cite{Arkani-Hamed:2010gh} through $n=10$ points. 
\section{Coplanar - intersecting cuts and path dependence}
An obvious degeneration of the above configurations would be to demand that all the lines lie in a plane and intersect each other. Here, we will see that the order in which the limit is taken determines the result. Recall that the form in ($\ref{eq:coplanarCut}$) is actually the form of the dual configuration in which all the dual lines are intersecting and we demand that they satisfy $\langle AB_\alpha ij \rangle >0, \forall i, j $. We can now take the limit $\langle A123\rangle = \langle AB_\alpha 12 \rangle = 0 $ or $\langle A123\rangle = \langle AB_\alpha 23 \rangle = 0$ which forces all the lines $AB_\alpha$ (which already intersect at $A$) to lie in the common plane $(123)$. We can perform a similar procedure on the form in $(\ref{eq:intersectingCut})$. We will show below that the results are significantly different.

First consider the intersecting cut. To make the configuration collapse to the plane $(123)$, we need a pole $\langle A123\rangle $ in addition to either $\langle AB_\alpha 12\rangle$ or $\langle AB_\alpha 23\rangle$. Note that there are two solutions to $\langle AB_\alpha 12 \rangle = \langle AB_\alpha 23 \rangle = 0$, one in which $(AB_\alpha)$ passes through $Z_2$ and the other in which it lies in the plane $(123)$. However, since all the regions in $B_\alpha$ are polygons, they are designed to have singularities only on their vertices. Thus the numerator is designed to kill the singularity in which the line lies in the plane $(123)$. This is precisely the singularity we are looking for. Hence we can achieve this limit only if the pole $\langle A123\rangle$ is present, which severely restricts the number of terms that can contribute to this cut. In fact, it is easy to see that only the triangles can contribute.  Thus, we are left with the result that at $L$ loops and $n$, points, if $A$ lies in the plane $(123)$, the corresponding region in $B_\alpha$ must be either the triangle $(12)(23)(34)$ or $(n1)(12)(23)$. 

We can derive the same result directly from the amplituhedron. Since we are interested in a configuration of coincident, coplanar lines in the MHV amplituhedron, we can parametrize them as follows
\beas
A = Z_1 + a_2 Z_2 + a_3 Z_3 \\
B_\alpha = Z_1 + b_\alpha Z_2
\eeas
and demand $\langle AB\bar{i}\bar{j}\rangle >0$. The mutual positivity is trivialized and the form is just the product of the form for each $B_\alpha$. It is not hard to see that the final result is
\bea
\label{eq:cc}
\Omega^{cc} = \frac{1}{a_2 a_3}\left(\prod_{\alpha = 1}^L \frac{1}{(a_2-b_\alpha )}+\prod_{\alpha = 1}^L \frac{1}{b_\alpha}\right).
\eea
The first term corresponds to the triangle $(n1)(12)(23)$ and the second to $(12)(23)(34)$.

In contrast with this simple result, the coplanar cut yields a far more complex residue. Indeed whenever $B_\alpha$ is in any triangle whose edge is either $(12)$ or $(23)$, the corresponding region in $A$ has the pole $\langle A123 \rangle$ required to collapse the configuration into the $(123)$ plane. 
\section{Moving beyond trivial mutual positivity}
\label{sec:nontrivial}
The results in equations (\ref{eq:coplanarCut}), (\ref{eq:intersectingCut}) and (\ref{eq:cc}) are valid for an arbitrary number of loops. While analytic all-loop results are few and far between, it is essential to realize that it was possible to obtain these results only because of the trivial mutual positivity condition. It is essentially equivalent to solving a one-loop problem. In this section, we begin exploring a few different configurations in which the mutual positivity conditions are not completely trivialized. We see that the associated geometries are far richer and the corresponding canonical forms more complex. In Section~\ref{sec:ladder}, we consider generalized ladder cuts where we cut only external propagators, while in Section~\ref{sec:furtherCuts} we examine several cuts which are directly related to the intersecting and coplanar cuts. 
\subsection{Ladder cuts}
\label{sec:ladder}
We consider the cut where our loops $(AB)_\alpha$, $\alpha = 1, \dots L$ all intersect one line, say $(12)$. Concretely, we are looking to find the form on the cut $\langle AB_\alpha 12\rangle = 0$. Let us write our form as 
\beas
\Omega = \prod_{\alpha = 1}^L \langle (AB)_\alpha \mathrm{d}^2A_\alpha \rangle \langle (AB)_\alpha \mathrm{d}^2B_\alpha \rangle \frac{1}{\langle (AB)_\alpha 12 \rangle} f[(AB)_\alpha].
\eeas
We expand $A_\alpha = Z_1 + x_\alpha Z_2 + z_\alpha Z_\star$ and take the residue $z_\alpha \rightarrow 0$ to obtain the ladder cut:
\bea
\label{eq: ladderform}
\Omega^{\text{ladder}} = \prod_{\alpha = 1}^L \mathrm{d}x_\alpha \langle (AB)_\alpha \mathrm{d}^2B_\alpha \rangle f[(AB)_\alpha]\Big\vert_{A_\alpha = 1+x_\alpha 2}.
\eea
We will determine $\Omega^{\text{ladder}}$ from the geometry of the amplituhedron. We can satisfy all but the mutual postivity condition by putting each loop $(AB)_\alpha$ in a Kermit
\bea
\label{eq:kermit}
&& A_\alpha = Z_1 + x_\alpha Z_2\\
&&\nonumber B_\alpha = -Z_1 + y_\alpha (Z_{i_\alpha} + w_\alpha Z_{i_\alpha +1}),
\eea
so that each cell is labelled by $L$ integers $\{i_1,\ldots,i_L\}$. Indeed, the conditions $\langle (AB)_\alpha ii+1 \rangle >0$ and the sign flip criterion are satisfied and each $(AB)_\alpha$ is in the one-loop amplituhedron so long as $x_\alpha , y_\alpha, w_\alpha >0$.  It remains to work out the implications of mutual positivity $\langle (AB)_\alpha (AB)_\beta \rangle >0$. Inserting (\ref{eq:kermit}) we find
\bea
\label{eq:ladderconditions}
&&\langle (AB)_\alpha (AB)_\beta \rangle = - \langle A_\alpha A_\beta B_\alpha B_\beta \rangle \\
&&\nonumber = y_\alpha y_\beta (x_\alpha - x_\beta) \left[ \langle 12i_\alpha i_\beta \rangle + w_\alpha \langle 12\,i_\alpha{+}1 i_\beta \rangle + w_\beta \langle 12\,i_\alpha\, i_\beta{+}1\rangle + w_\alpha w_\beta \langle 12\,i_\alpha {+}1\, i_\beta {+}1\rangle \right].
\eea
Depending on the relative positions of $\alpha$ and $\beta$, we have the following cases:
\begin{itemize}
    \item $i_{\alpha} < i_{\alpha + 1} < i_{\beta} < i_{\beta + 1}$ \\
\end{itemize}
In this case, we have $\langle 12 i_\alpha i_\beta \rangle > 0\,,\, \langle 12 i_{\alpha+1}\, i_\beta \rangle > 0\, , \, \langle 12 i_\alpha \,i_{\beta +1} \rangle > 0\,$ and\\  $\langle 12 i_{\alpha + 1}\, i_{\beta + 1}\, \rangle > 0$. Hence (\ref{eq:ladderconditions}) reduces to
\bea
\label{eq:lc1}
\left( x_\alpha - x_\beta \right) > 0. 
\eea

\begin{itemize}
    \item $i_{\alpha} < i_{\alpha + 1} = i_{\beta} < i_{\beta + 1}$ \\
\end{itemize}
In this case, $\langle 12 i_\alpha i_\beta \rangle > 0\,,\, \langle 12 i_{\alpha+1}\, i_\beta \rangle = 0\, , \, \langle 12 i_\alpha \,i_{\beta +1} \rangle > 0\,$ and\\  $\langle 12 i_{\alpha + 1}\, i_{\beta + 1}\, \rangle > 0$ and (\ref{eq:ladderconditions}) again reduces to
\bea
\label{eq:lc2}
\left( x_\alpha - x_\beta \right) > 0.
\eea

\begin{itemize}
    \item $i_{\alpha} = i_{\beta} < i_{\alpha + 1} = i_{\beta + 1}$ \\
\end{itemize}
This configuration makes (\ref{eq:ladderconditions}) collapse to
\bea
\label{eq:lc3}
\left( w_\beta - w_\alpha \right) \left( x_\alpha - x_\beta\right) > 0.
\eea
At $L$ loops, we will have $3L$ variables $x_\alpha, y_\alpha, w_\alpha$ satisfying the inequalities above. Let us denote by $g_{\lbrace i_1 \dots i_L\rbrace}(x_\alpha, w_\alpha)$ the canonical form associated with the $L$-loop configuration. Note that the $y_\alpha$ factor out of the problem since they are unconstrained variables. We can write
\bea
\Omega^{\text{ladder}} = \sum_{\lbrace i_1 \dots i_L\rbrace} \prod_{\alpha}\frac{\mathrm{d}y_\alpha}{y_\alpha}\, \mathrm{d}x_\alpha\, \mathrm{d}w_\alpha\, g_{\lbrace i_1 \dots i_L\rbrace }\,(x_\alpha, w_\alpha), 
\eea
where $\sum_{\lbrace i_1 \dots i_L \rbrace}$ stands for a sum over all configurations at $L$ loops. To compute the canonical form for this space, we need to triangulate it. However, in order to add the canonical forms associated with different pieces in the triangulation, we need to write the form of each piece in a coordinate invariant way. The variables $x_\alpha$ are the same for all cells but the $y_\alpha$ and $w_\alpha$ are cell dependent. We can obtain coordinate invariant expressions by noting that the point of intersection of the line $(AB)$ with the plane $(1Z_i Z_{i+1})$ is by the Schouten identity
\bea
(AB) \cap (1 Z_i Z_{i+1}) = \langle ABii+1 \rangle Z_1 - \langle AB1i+1 \rangle Z_i + \langle AB1i \rangle Z_{i+1}.
\eea
Comparing with (\ref{eq:kermit}), we read off
\bea
\label{eq: yw}
y = \frac{\langle AB1i{+}1 \rangle}{\langle ABii{+}1 \rangle} \qquad y \,w = - \frac{\langle AB1i \rangle}{\langle ABii{+}1 \rangle} \implies w = -\frac{\langle AB1i \rangle}{\langle AB1i{+}1 \rangle}.
\eea
From the measure associated with the Kermit, we have
\beas
\frac{\mathrm{d}y\, \mathrm{d}w}{y\, w} = \frac{\langle AB\mathrm{d}^2B \rangle \langle A1ii{+}1\rangle^2}{\langle AB1i\rangle \langle AB1i{+}1\rangle \langle ABii+1\rangle} \implies \frac{\mathrm{d}y\, \mathrm{d}w}{y } = -\frac{\langle AB\mathrm{d}^2B \rangle \langle A1ii{+}1\rangle^2}{\langle AB1i{+}1\rangle ^2 \langle ABii{+}1\rangle}.
\eeas
With this, we can write $\Omega^{\text{ladder}}$ in an invariant way as
\begin{equation}
\begin{split}
\Omega^{\text{ladder}} =&(-1)^L \sum_{\lbrace i_1 \dots i_L\rbrace} \prod_{\alpha}\frac{\mathrm{d}x_\alpha\langle A_\alpha 1i_\alpha i_\alpha{+}1\rangle^2}{\langle (AB)_\alpha 1i_\alpha {+}1 \rangle^2 \langle (AB)_\alpha i_\alpha i_\alpha{+}1\rangle}\\&\times g_{\lbrace i_1 \dots i_L\rbrace }\left(x_\alpha, -\frac{\langle (AB)_\alpha 1i_\alpha \rangle}{\langle (AB)_\alpha 1i_\alpha {+}1\rangle}\right).
\end{split}
\end{equation}
Let us work out a few examples at low loop orders to get a better idea of how to write the form explicitly. The first case $L = 1$ is trivial, since $g(x,w) = 1/(xw)$ and we have
\begin{equation}
\begin{split}
\Omega^{\text{ladder}} =& -\sum_i \frac{\langle A_1ii{+}1\rangle^2}{\langle AB1i{+}1 \rangle^2 \langle ABii{+}1 \rangle }\frac{1}{x}\frac{-1}{\frac{\langle AB1i\rangle}{\langle AB1i{+}1\rangle}} \\ =& \frac{1}{x}\sum_i \frac{\langle A1ii{+}1\rangle^2}{\langle AB1i \rangle \langle AB1i{+}1 \rangle \langle ABii{+}1\rangle}.
\end{split}
\end{equation}
At two loops, the function $g_{\lbrace i_1, i_2 \rbrace}$ is
\beas
&&\frac{1}{w_1 w_2 x_2(x_1-x_2)}, \qquad\qquad \text{if  } i_1<i_2,\\
&&\frac{1}{w_1 w_2 x_1(x_1-x_2)}, \qquad\qquad \text{if  } i_1>i_2,\\
&&\frac{1}{x_2(x_1-x_2)}\frac{1}{w_1 (w_2 -w_1)} +\frac{1}{x_1(x_2-x_1)}\frac{1}{w_2 (w_1 -w_2)}, \qquad\qquad \text{if  } i_1=i_2.
\eeas
Moving to three loops, for the set $\{i_1,i_2,i_3\}$ there are three possibilities: (i) all three indices are distinct, (ii) Two of the indices are equal, or (iii) all three indices are equal. For each possibility, the indices can be ordered in a variety of ways. Furthermore, in the degenerate cases we must break these orderings into smaller pieces in order to triangulate the space. For example, if $i_1=i_2$ we must consider both cases $x_{i_1}<x_{i_2}$ and $x_{i_1}>x_{i_2}$ separately. Repeating this for an arbitrary number of loops it is easy to see that one possible triangulation which covers all possibilities exactly once is given by specifying the following:
\begin{itemize}
    \item A partition $N = \lbrace N_1, \dots N_m \rbrace$ of $L$ i.e., $\sum_i N_i = L$ and $N_i \geq 1$ along with an associated set of integers $\mathcal{J}_N = \lbrace j_1 , \dots j_m \rbrace $ of equal length such that $3 \leq j_1 < j_2 < \dots \leq n-1$. The integer $N_i$ represents how many of the loops $(AB)_\alpha$ are in the Kermit labelled by $\left[1\,2\,3; 1\,j_i\,j_{i+1}\right]$
    \item A permutation $\Pi = \lbrace \pi_1 \dots \pi_L \rbrace$ of $\lbrace 1, \dots L\rbrace$.
\end{itemize}
The sum over all the cells is carried out by summing over all possible $\mathcal{N}, J_{\mathcal{N}}, \Pi$. For the sake of compactness, we define another quantity
\bea
\label{eq:W}
W[s,e,\Pi,j] \equiv \frac{1}{w_{\pi_s}(w_{\pi_{s+1}}-w_{\pi_s}) \cdots (w_{\pi_{e}}-w_{\pi_{e-1}})}\Big\vert_{ w_{\pi_{\alpha}}=\frac{-\langle (AB)_{\pi_{\alpha}} 1j\rangle}{\langle (AB)_{\pi_{\alpha}}1j{+}1\rangle}}.
\eea
In terms of this function, the form for the ladder cut can be written as
\begin{equation}
\begin{split}
\label{final}
\Omega^{\text{ladder}}=& (-1)^L \sum_{N,J_{N},\Pi} \frac{1}{x_{\pi_L} (x_{\pi_{L-1}}-x_{\pi_L}) \dots (x_{\pi_{1}}-x_{\pi_2})} \\
&\times\left(\prod_{\alpha=1}^{N_1} \frac{\langle A_{\pi_\alpha}1j_1j_1{+}1\rangle^2}{\langle (AB)_{\pi_\alpha}1j_1{+}1\rangle^2 \langle (AB)_{\pi_\alpha}j_1j_1{+}1\rangle}\right) W[1,N_1,\pi,j_1]  \\
&\times \left(\prod_{\alpha = 1{+}N_1}^{N_2} \frac{\langle A_{\pi_\alpha}1j_2j_2{+}1\rangle^2}{\langle (AB)_{\pi_\alpha}1j_2{+}1\rangle^2 \langle (AB)_{\pi_\alpha}j_2j_2{+}1\rangle}\right) W[1{+}N_1,N_1{+}N_2,\pi,j_2] \\
&\times\vdots\\
&\times\left(\prod_{\alpha=1{+}N_1{+}\cdots{+} N_{m-1}}^{N_m} \frac{\langle A_{\pi_\alpha}1j_mj_m{+}1\rangle^2}{\langle (AB)_{\pi_\alpha}1j_m{+}1\rangle^2 \langle (AB)_{\pi_\alpha}j_mj_m{+}1\rangle}\right)\\
&\times W[1{+}N_1{+}\cdots+N_{m-1},N_m,\pi,j_m].
\end{split}
\end{equation}
\subsection{Extra free lines}
\label{sec:furtherCuts}
In this section, we will consider a series of cuts in which the configuration of lines $(AB)_\alpha$ are minor modifications to the coplanar and collinear cut. In each case, we consider an extra line which allows for non trivial mutual positivity. In order of increasing difficulty, some of the types of cuts we consider involve the following configurations of lines:
\begin{itemize}
\item {\underline{{\bf Cut 1}}: } $L{-}1$ loops intersecting in a common point $A$, with each line passing through one of the external $Z_i$. We can denote these lines as $Ai$. An additional line passes through some $Z_j$, but does not intersect the lines $Ai$ in $A$. Denoting this line by $Bj$, the non trivial mutual positivity conditions are $\langle AiBj \rangle > 0$.
\item  {\underline{{\bf Cut 2}}: } $L{-}1$ loops $Ai$, intersecting in a common point $A$ and passing through some $Z_i$ with the $L^{\text{th}}$ line $CD$ completely free. Here, the addtional constraint is $\langle CDAi \rangle >0$.
\item  {\underline{{\bf Cut 3}}: }$L{-}1$ loops $AB_\alpha$ which intersect at $A$ with the $L^{\text{th}}$ line $CD$ intersecting two of the lines $AB_i$ and $AB_j$ resulting in the non trivial constraint $\langle AB_\alpha CD \rangle >0$ with $\alpha \neq i,j$.
\item  {\underline{{\bf Cut 4}}: } $L{-}1$ loops intersecting in a common point $A$ with the $L^{\text{th}}$ line completely free. This is a generalization of the above cut.
\end{itemize}
The first two cuts are generalizations of the $(4L{-}4)$-cuts of Section~\ref{sec:cyclicpolytopecuts} while the next two are related to the $(2L{-}4)$ cuts of Section~\ref{sec:deepcuts}.\\

\begin{figure}[t]
    \centering
    \includegraphics[scale=0.4]{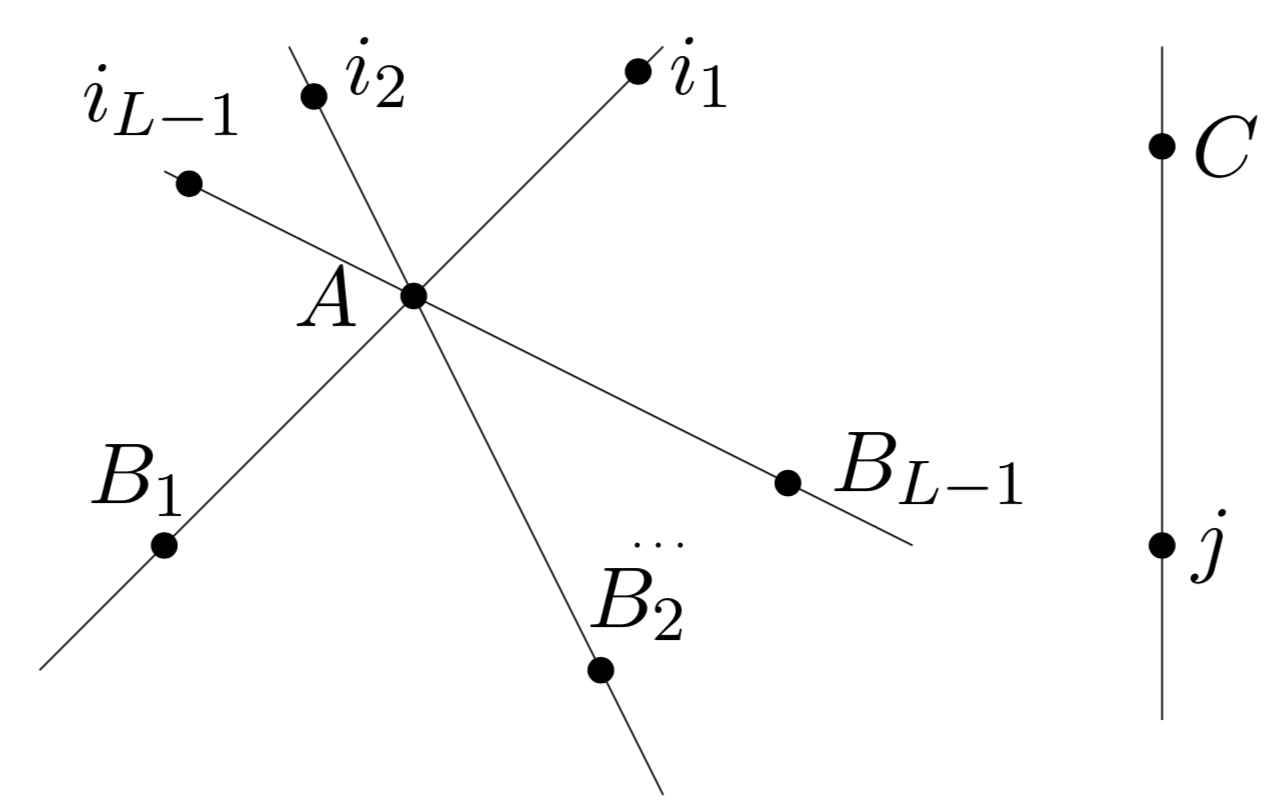}
    \caption{Cut 1, where $L{-}1$ loops intersect in a common point $A$ as well as $L{-}1$ points $Z_{i_1},\ldots,Z_{i_{L-1}}$, and the $L^{\text{th}}$ loop intersects an additional point $Z_j$.}
    \label{fig:cut1}
\end{figure}

\noindent {\underline{{\bf Cut 1}}: }\\
Here, the configuration of lines $Ai$ is the same as in Section \ref{sec:cyclicpolytopecuts}, with modifications for the $L^{\text{th}}$ loop as shown in Figure~\ref{fig:cut1}.
We begin by solving this problem at four and five points to illustrate the complications presented by mutual positivity.

A generic configuration at four points includes $L_1$ lines passing through $Z_1$, $L_2$ lines passing through $Z_2$, $L_3$ through $Z_3$ and $L_4$ through $Z_4$. This cut has already computed in \cite{IntoTheAmplituhedron} using a slightly different approach. Here, we will merely present a simple example of a three-loop cut with lines $A1$ and $A2$ intersecting at $A$ and passing through $Z_1$ and $Z_2$ respectively.The third loop $(B3)$ passes through $Z_3$ but is otherwise unconstrained. We can be parametrize the points $A$ and $B$ as
\begin{equation}
A=Z_1-a_2Z_2-a_3Z_3-a_4Z_4,\qquad B=Z_1-b_2Z_2+b_3Z_3-b_4Z_4.
\end{equation}
The constraints $\langle A1\bar{i}\bar{j}\rangle>0$,$\langle A2\bar{i}\bar{j}\rangle>0$ and $\langle B3\bar{i}\bar{j}\rangle>0$ are trivially satisfied by $a_i>0$ and $b_2,b_4>0$. We are left with the mutual positivity conditions 
\begin{equation}
\begin{split}
-\langle A1B3\rangle&=(a_4b_2-a_2b_4)\langle1234\rangle<0,\\
-\langle A2B3\rangle&=(a_4-b_4)\langle1234\rangle<0.
\end{split}
\end{equation}
The canonical form associated to these inequalities is trivial to obtain: 
\begin{equation}
\Omega=\frac{\langle 1234\rangle^4\langle B123\rangle\langle A\mathrm{d}^3 A\rangle\langle B\mathrm{d}^3B\rangle}{\langle A123\rangle\langle A124\rangle\langle AB23\rangle\langle AB13\rangle\langle B124\rangle\langle B134\rangle\langle B234\rangle},    
\end{equation}
which matches the three-loop integrand evaluated on the same cut and agrees with the general result for the corner cut in \cite{IntoTheAmplituhedron}. Note the presence of the poles $\langle AB13 \rangle$ and $\langle AB23 \rangle$ is due to the mutual positivity constraint. This demonstrates that this condition is introducing new physical boundaries into the geometry.

Moving on to five points, we begin with $L=3$. Consider the configuration of the cyclic polytope cut of Section \ref{sec:cyclicpolytopecuts}, where we have lines $A1$ and $A2$ which intersect at $A$ and additionally pass through $Z_1$ and $Z_2$. The third loop $(AB)_3=(1B)$ passes through $Z_1$ but does not intersect the other lines in $A$. The point $B$ has two degrees of freedom since it is constrained to lie on the line $(1B)$. By imposing the inequalities
\begin{equation}
\langle A\alpha\bar{i}\bar{j}\rangle>0,\quad \alpha=1,2,\qquad \langle 1B\bar{i}\bar{j}\rangle>0,\qquad \langle A21B\rangle>0,
\end{equation}
on the points $A$ and $B$, the associated canonical form is
\begin{equation}
\begin{split}
\Omega_5^{(3)}\vert_{\text{cut 1}}=&\frac{-\langle A\mathrm{d}^3A\rangle \langle 1B\mathrm{d}^2B\rangle}{\langle A145\rangle \langle A134\rangle \langle B145\rangle \langle B125\rangle \langle AB12\rangle \langle A345\rangle \langle A123\rangle \langle B134\rangle}\\ \times& \big(\langle A123\rangle \langle B134\rangle \langle 1245\rangle^2\langle 1345\rangle^2 + \langle A145\rangle \langle B145\rangle \langle 1234\rangle^2 \langle 1235\rangle^2\\&+\langle A145\rangle \langle B123\rangle \langle 1345\rangle^2 \langle 1235\rangle \langle 1245\rangle \big).
\end{split}
\end{equation}
Next consider the corresponding $L=4$ configuration where the first three loops are $A\alpha$ for $\alpha = 1,2,3$, and the fourth line is $(AB)_4=(1B)$. As we found in Section~\ref{sec:cyclicpolytopecuts}, the point $A$ must be in the tetrahedron with vertices $Z_3,Z_4,Z_5,-Z_1$. Here we can parametrize the two points $A$ and $B$ as
\begin{equation} 
A = Z_1+ a_3 Z_3 + a_4 Z_4 + a_5 Z_5 \qquad B = -Z_2 + b_3 Z_3+ b_4 Z_4.
\end{equation}
Demanding that the inequalities 
\begin{equation}
\langle A\alpha\bar{i}\bar{j}\rangle>0,\qquad \langle 1B\bar{i}\bar{j}\rangle>0,\qquad \langle AB\alpha1\rangle>0,\quad \alpha=1,2,3,
\end{equation}
are satisfied, we find the associated canonical form
\begin{equation}
\begin{split}
\Omega_5^{(4)}\vert_{\text{cut 1}}=&\frac{\langle A\mathrm{d}^3A\rangle \langle 1B\mathrm{d}^2B\rangle}{\langle A345 \rangle \langle A145\rangle \langle A135\rangle \langle A134\rangle\langle B145\rangle \langle B125\rangle \langle B134\rangle\langle AB12\rangle \langle AB13\rangle}\\ \times&
\big( \langle A135\rangle^2\langle B134\rangle^2 \langle 1245\rangle^2 \langle 1345\rangle - \langle A135\rangle\langle A145\rangle \langle B123\rangle\langle B134\rangle  \langle 1245\rangle \langle 1345\rangle^2\\ &-\langle A134\rangle \langle A135\rangle \langle B135\rangle \langle B134\rangle \langle 1245\rangle^2 \langle 1345\rangle+\langle A134\rangle^2 \langle B125\rangle^2 \langle 1345\rangle^3\\ &+ \langle A123\rangle \langle A145\rangle \langle B145\rangle \langle B134\rangle \langle 1235\rangle \langle 1345\rangle^2 \\ &- \langle A134\rangle \langle A135\rangle \langle B124\rangle \langle B125\rangle \langle 1345\rangle^3 \big).
\end{split}
\end{equation}
In both these cases, we can see the poles due to mutual positivity. The all-loop extension of this configuration with $L-1$ lines $A1, A2, \ldots A(L{-}1)$ passing through $1, \ldots (L{-}1)$, respectively, and the $L^{\text{th}}$ line $(AB)_L=(1B)$ passing through $Z_1$ but not $A$ can be similarly obtained on a case-by-case basis. However, we do not yet have an analytic expression valid for all $L$.\\

\begin{figure}[t]
    \centering
    \includegraphics[scale=0.5]{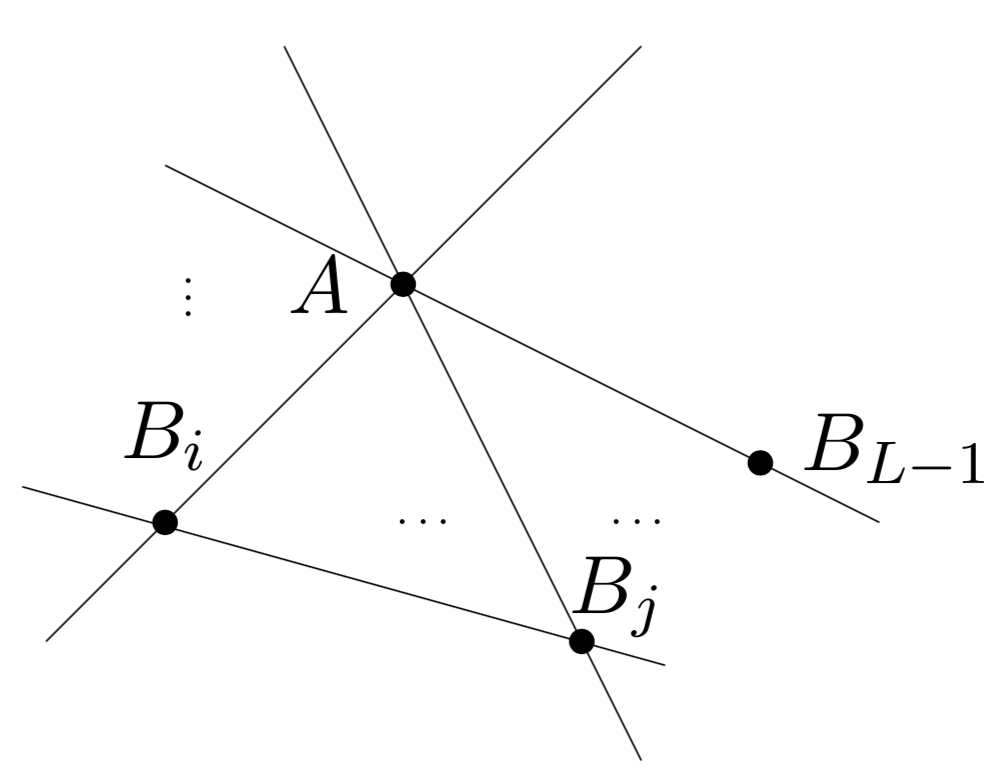}
    \caption{Cut 3 where $L{-}1$ loops intersect in a common point $A$, and the $L^{\text{th}}$ loop intersects lines $i$ and $j$.}
    \label{fig:cut3}
\end{figure}

\noindent {\underline{{\bf Cut 2}}: }\\
We now lift the constraint that the extra line passes through one of the external points. However, we will still consider the configuration where $L-1$ lines $A1, A2, \ldots A(L{-}1)$ pass through $1, \ldots (L{-}1)$, respectively, so the configuration is identical to that of Fig.~\ref{fig:cut1} with the line $(Cj)\rightarrow(CD)$. The relevant inequalities are
\bea
\langle CD \bar{i}\bar{j}\rangle \,>\,0  \qquad \langle AkCD\rangle \,>\,0 \qquad \langle Ak\bar{i}\bar{j}\rangle >0 \qquad k = 1, \dots L.
\eea
We parametrize $CD$ by putting it in the Kermit:
\bea
C = Z_1 + \alpha_1 Z_a +\alpha_2 Z_{a+1},\\
D = -Z_1 + \beta_1 Z_b +\beta_2 Z_{b+1}.
\eea
The one-loop constraints $\langle CD\bar{i}\bar{j}\rangle>0$ enforce positivity of $\alpha_i$ and $\beta_i$. As before, the one-loop conditions on the lines $Ai$, which are independent of the line $CD$ imply that $A$ must lie in the cyclic polytope $\mathrm{Conv}\,[L,L+1,..-1]$. The mutual positivity conditions reduce to a single condition,
\begin{equation}
\begin{split}
\langle AkCD\rangle = &\langle Ak1b \rangle \beta_1 + \langle Ak1b{+}1 \rangle \beta_2 + \langle Ak1a \rangle \alpha_1 + \langle Akab \rangle \alpha_1\beta_1 +   \langle Akab{+}1 \rangle \alpha_1\beta_2 \\
&+\langle Ak1a{+}1 \rangle \alpha_2 + \langle Aka{+}1b \rangle \alpha_2\beta_1 + \langle Aka{+}1b{+}1 \rangle \alpha_2\beta_2. 
\end{split}
\end{equation}
Although we do not have a complete understanding of this system of inequalities, in some simple cases an analytical solution is possible. For example, for $n=4$ the free loop line $CD$ is in the Kermit $\left[ 123;134\right]$, and the form is given by
\begin{equation}
\begin{split}
\Omega_4^{(4)}\vert_{\text{cut}}=&\frac{\langle A\mathrm{d}^3A \rangle \langle CD\mathrm{d}^2C\rangle \langle CD\mathrm{d}^2 D \rangle \langle 1234\rangle ^3}{\langle A123\rangle \langle A234\rangle \langle A134\rangle \langle A124\rangle \langle CD14\rangle \langle CD23\rangle \langle CD34\rangle \langle CDA2\rangle \langle CDA1\rangle}\\
\times&\big( -\langle CD34\rangle \langle A123\rangle \langle A124\rangle + \langle A123\rangle \langle A234\rangle \langle CD14\rangle \\
&+\langle A134\rangle \langle CD12\rangle \langle A234\rangle - \langle A134\rangle \langle A124\rangle \langle CD23\rangle \big).
\end{split}
\end{equation}
\noindent {\underline{{\bf Cuts 3 and 4}}: }\\
Finally, we can also consider cuts which relax conditions on the  $2L-4$ cuts discussed in Section~\ref{sec:deepestcuts}. For example, we can consider $L-1$ loops intersecting in a common point $A$ (but not passing through any external $Z_i$), and the $L^{\text{th}}$ line intersecting two of the loops $(AB)_i$ and $(AB)_j$, as pictured in Figure~\ref{fig:cut3}.
The $L=3$ configuration is simply the coplanar cut discussed above, but $L=4$ is more interesting. Here we can take the first three loop lines to intersect, and the fourth line to cut $(AB)_1$ and $(AB)_2$. We can write $(AB)_\alpha=AB_\alpha$ for $\alpha=1,2,3$ and for the fourth loop $(AB)_4=(B_1B_2)$. The inequalities defining the four-loop amplituhedron become
\begin{equation}
\langle AB_\alpha \bar{i}\bar{j}\rangle>0,\qquad \langle B_1B_2\bar{i}\bar{j}\rangle>0,\qquad \langle AB_1B_2B_3\rangle>0,  
\end{equation}
where there is only a single remaining mutual positivity condition. Parametrizing the intersection point $A$ and the points $B_1,B_2,B_3$ as in Section~\ref{sec:deepcuts},
\begin{equation}
\begin{split}
A&=Z_1+a_2Z_2+a_3Z_3+a_4Z_4,\qquad B_\alpha=Z_1+x_\alpha Z_2+y_\alpha Z_3,\quad \alpha=1,3\\
B_2&=Z_1+x_2Z_3+y_2Z_4,
\end{split}
\end{equation}
(where we choose a different parametrization for $B_2$ so the configuration is not too degenerate) we get several quadratic inequalities and a single cubic inequality,
\begin{equation}
\begin{split}
\langle AB_1B_2B_3\rangle=&\left[x_1y_3+x_3(x_2-y_1)-x_1x_2\right]a_4+\left(x_1y_2-x_2x_3\right)a_3+\left(x_2y_3-y_1y_2\right)a_2\\&+y_1y_2x_3-x_1y_2y_3>0.
\end{split}
\end{equation}
Completing the triangulation we get for the canonical form
\begin{equation}
\Omega_4^{(L)}\vert_{\text{cut}}=\frac{\langle A\mathrm{d}^3 A\rangle\prod\limits_{\alpha=1,3}\langle AB_\alpha\mathrm{d}^2 B_\alpha\rangle\langle B_1B_2\mathrm{d}^2B_1\rangle\langle B_1B_2\mathrm{d}^2 B_2\rangle N\left(AB_\alpha,B_1B_2\right)}{\prod\limits_{\alpha=1,3}\prod\limits_{i=1}^4\left(\langle AB_\alpha ii{+}1\rangle\right)\prod\limits_{i=1}^4\left( \langle B_1B_2ii{+}1\rangle\right)\langle AB_1B_2B_3\rangle},
\end{equation}
where the numerator $N\left(AB_\alpha,B_1B_2\right)$ is a sum with several hundred terms. We have verified the result of this calculation matches the full four-point four-loop integrand, which is a sum of eight local diagrams, symmetrized over all loop momenta and cyclically summed over external legs and given explicitly in momentum twistor variables in \cite{ArkaniHamed2015}, evaluated on this cut.

The same cut at five points is also solvable with the amplituhedron, although we have yet to find a particularly simple representation of the canonical form which suggests a generalization to higher points and loops. We plan to revisit these problems as well as generalized corner cuts in future work.
\section{Conclusion}
The all-loop amplituhedron is a remarkable mathematical object capturing the complicated loop-level structure of scattering amplitudes in planar $\mathcal{N}=4$ sYM in geometric form directly in the physical kinematic space. This paper has been concerned with the practical application of this geometric picture to make predictions about the MHV loop integrand, valid for any number of particles and any number of loops, which are completely hidden in the usual unitarity or recursion-based methods. In particular we studied a series of cuts which probed the part of the loop-integrand which is, in the Feynman diagram expansion, encoded in the subset of diagrams with many internal propagators which have complicated branch-cut structure. We found remarkably simple expressions for the canonical forms for these ``maximally intersecting'' cuts. The topological winding formulation of the amplituhedron of~\cite{Arkani-Hamed:2017tmz} was crucial in deriving our results. In fact without this sign flip picture even a qualitative description of the canonical forms (\ref{eq:intersectingCut}) and (\ref{eq:coplanarCut}), the central results of this paper, would likely be impossible. However, from the perspective of the amplituhedron, the factorization of the canonical forms on the intersecting and coplanar cuts is completely trivial and follows directly from the definition of the geometry. However, our analysis reveals an even greater simplicity than one would na\"{i}vely guess: for the intersecting cut the allowable space for the intersection point is naturally triangulated by a simple collection of tetrahedra, while the remaining degrees of freedom of the loop lines live inside a polygon.

This work is a continuation of a systematic exploration of the facets of the amplituhedron for all $n,k,L$. As such, there are a number of avenues for further investigation: first, there are the unfinished cuts presented in Section~\ref{sec:furtherCuts} which gradually relax some of the constraints imposed on the maximally intersecting cuts we solved. The most interesting (and complicated) extension of the all-loop results presented here involve $L{-}1$ lines intersecting in a common point $A$ with the $L^{\text{th}}$ line free; solving this cut would amount to a complete understanding of the MHV two-loop geometry. Although the direct product form of the solutions obtained to all-loop orders will of course not remain, preliminary considerations suggest that simple geometrical decompositions of the canonical forms do persist to these more generic cuts. Another natural starting point for further work is to consider the same maximally intersecting cuts for $k\geq1$ i.e., different helicity sectors. For example, by parity conjugation the NMHV five-point coplanar cut is simply the $R$-invariant $[12345]$ multiplying the result derived in this paper at five points. In the general $n,k$ case although the product form will remain, the sign flip conditions change for both the external data and the loop momentum variables; however, it is likely that just as in the MHV configuration considered here, these problems will ultimately reduce to finding the right way of understanding the corresponding one-loop geometries for arbitrary $n,k$. Finally, there is another class of facets of the amplituhedron which are of physical interest. These involve unitarity cuts which trivialize the inequalities involving external data while leaving the mutual positivity conditions untouched. An example of these are the ``corner cuts'' computed in~\cite{IntoTheAmplituhedron} at four points where loop lines pass through either $Z_i$ or $(i{-}1ii{+}1)$. A detailed understanding of such corner cuts, along with complete knowledge of the structure of the integrand on the maximally intersecting cuts initiated here, would be invaluable to the goal of reconstructing the full loop integrand directly from the amplituhedron. 

\acknowledgments
This work was made possible with the support of Nima Arkani-Hamed and Jaroslav Trnka. C. L. is supported in part by DOE grant No. DE-SC0009999 and the funds of the University of California.
\appendix
 \section{Constraints from projecting positive data}
 \label{sec:constraints}
In this appendix, we derive the constraints that result from projecting positive data. We derive the constraints that must be satisfied by 3 dimensional data which are the result of projecting four dimensional positive data. Let us start with $n=5$. We can add one extra component and turn them into 4D data. 
\beas
Z_i = \begin{pmatrix}
{\bf z}_i\\
c_i 
\end{pmatrix} \qquad i=1,..5
\eeas
The 3D ${\bf z}_i$ can be thought of as coming from positive 4D data if we can add a fourth component such that the resulting 4D data are positive. Thus at 5 points, we need to demand 
\beas
\langle 1234 \rangle >0, \langle 1235 \rangle >0, \langle 1245 \rangle >0, \langle 1345 \rangle >0, \langle 2345 \rangle >0  
\eeas
The resulting system of equations can be written in the following way.
\bea
\begin{pmatrix}
-\langle {\bf 234} \rangle && \langle {\bf 134} \rangle && -\langle {\bf 124} \rangle && \langle {\bf 123} \rangle && 0\\
-\langle {\bf 235} \rangle && \langle {\bf 135} \rangle && -\langle {\bf 125} \rangle && 0 &&\langle {\bf 123} \rangle \\
-\langle {\bf 245} \rangle && \langle {\bf 145} \rangle && 0 &&  -\langle {\bf 125} \rangle && \langle {\bf 124} \rangle \\
-\langle {\bf 345} \rangle && 0 && \langle {\bf 145} \rangle && -\langle {\bf 135} \rangle && \langle {\bf 134} \rangle \\
0 && -\langle {\bf 345} \rangle && \langle {\bf 245} \rangle && -\langle {\bf 235} \rangle && \langle {\bf 345} \rangle \\
\end{pmatrix}  \begin{pmatrix}
 c_1\\ c_2\\ c_3\\ c_4\\ c_5 
\end{pmatrix} = A^T c\, > \, 0
\eea
Thus, we can think of the 3D data, ${\bf z}_i$ as coming from 4D positive data if this system of inequalities as a solution,. The condition for the existence of a solution for a system of linear inequalities is given by Gordan's theorem which states,

\begin{thm}
Exactly one of the following systems has a solution.\\
(1) $y^T A > 0 $ for some $y \in \mathbf{R}^n$\\
(2) $A x = 0, x\geq 0$ for some non zero $x \in \mathbf{R}^n$
\end{thm}
Thus the condition for the existence of a solution to our system is that the null vectors cannot have all positive entries. To find the null eigenvectors of $A$, we first note that the Schouten identity in three dimensions is
\bea
\langle {\bf 123}\rangle {\bf 4}-\langle {\bf 234}\rangle {\bf 1}+\langle {\bf 341}\rangle {\bf 2}-\langle {\bf 412}\rangle {\bf 3}=0
\eea 
\bea
A =
\begin{pmatrix}
-\langle {\bf 234} \rangle && -\langle {\bf 235} \rangle  && -\langle {\bf 245} \rangle && -\langle {\bf 345} \rangle  && 0\\
\langle {\bf 134} \rangle && \langle {\bf 135} \rangle && \langle {\bf 145} \rangle && 0 && - \langle {\bf 345} \rangle \\
-\langle {\bf 124} \rangle &&-\langle {\bf 125} \rangle && 0 &&\langle {\bf 145} \rangle &&\langle {\bf 135} \rangle\\
\langle {\bf 123} \rangle && 0 &&-\langle {\bf 125} \rangle &&-\langle {\bf 135} \rangle &&-\langle {\bf 235} \rangle \\
0 && \langle {\bf 123}\rangle && \langle {\bf 124}\rangle && \langle {\bf 134}\rangle&& \langle {\bf 345}\rangle 
\end{pmatrix}
\eea
We can easily see that any vector of the form $\begin{pmatrix}
\langle {\bf 5ab}\rangle && -\langle {\bf 4ab}\rangle && \langle {\bf 3ab}\rangle && -\langle {\bf 2ab}\rangle && \langle {\bf 1ab}\rangle
\end{pmatrix}$ is a null eigenvector as a consequence of the Schouten identity. Here ${\bf a}$ and ${\bf b}$ are any two 3D vectors.
From Gordan's theorem, the condition for the existence of a solution and consequently the constraint on the 3D data is that not all entries of the null vector are positive. Let us choose ${\bf a}={\bf z}_1$ and ${\bf b} = {\bf z}_2$. Then $\lbrace\langle {\bf 512}\rangle,-\langle {\bf 412}\rangle ,\langle {\bf 312}\rangle\rbrace$ aren't all positive or the sequence $\lbrace\langle {\bf 125}\rangle,\langle {\bf 124}\rangle ,\langle {\bf 123}\rangle\rbrace$ has less than 2 sign flips. However, in this case we cannot say anything about the sign flips of the sequences resulting from a different choice of ${\bf a}$ and ${\bf b}$. Furthermore, any one of them having the wrong flip pattern is sufficient to show that this 3D data cannot arise from positive 4D data. \\

This can be easily extended beyond $n=5$. At an arbitrary $n$, we have to impose positivity of all ordered minors $\langle ijkl \rangle$ with $i<j<k<l$. This results in a similar system of inequalities with null eigenvectors of the form 
\bea
\lbrace (-1)^i \langle {\bf (n-i)ab} \rangle  \rbrace , \qquad i=1,2...n-1
\eea 
which leads to a similar constraint on the signs. 

\bibliographystyle{my-JHEP}
\bibliography{LoopCutRefs}

\providecommand{\url}[1]{#1}\providecommand{\href}[2]{#2}\begingroup\raggedright\begin{thebibliography}{10}

\bibitem{Arkani-Hamed:2018caj}
N.~Arkani-Hamed, C.~Langer, A.Y.~Srikant and J.~Trnka, \emph{\textit{Deep into
  the Amplituhedron: Amplitude Singularities at All Loops and Legs}},
  \href{https://doi.org/10.1103/PhysRevLett.122.051601}{\emph{Phys. Rev. Lett.}
  {\bfseries \textbf{122}} (2019)}
  [\href{https://arxiv.org/abs/1810.08208}{{\ttfamily arXiv:1810.08208}}].

\bibitem{TheAmplituhedron}
N.~Arkani-Hamed and J.~Trnka, \emph{\textit{The Amplituhedron}},
  \href{https://doi.org/10.1007/JHEP10(2014)030}{\emph{JHEP} {\bfseries
  \textbf{1410}} (2014)} [\href{https://arxiv.org/abs/1312.2007}{{\ttfamily
  arXiv:1312.2007}}].

\bibitem{positiveGrass}
N.~Arkani-Hamed, J.L.~Bourjaily, F.~Cachazo, A.~Goncharov, A.~Postnikov and
  J.~Trnka, \emph{{Scattering Amplitudes and the Positive Grassmannian}}.

\bibitem{hodges}
A.~Hodges, \emph{{Eliminating spurious poles from gauge-theoretic amplitudes}},
  \href{https://doi.org/10.1007/JHEP05(2013)135}{\emph{JHEP} {\bfseries 05}
  (2013) 135} [\href{https://arxiv.org/abs/0905.1473}{{\ttfamily
  arXiv:0905.1473}}].

\bibitem{Arkani-Hamed:2017tmz}
N.~Arkani-Hamed, Y.~Bai and T.~Lam, \emph{{Positive Geometries and Canonical
  Forms}}, \href{https://doi.org/10.1007/JHEP11(2017)039}{\emph{JHEP}
  {\bfseries 11} (2017) 039}
  [\href{https://arxiv.org/abs/1703.04541}{{\ttfamily arXiv:1703.04541}}].

\bibitem{Bern:2014kca}
Z.~Bern, E.~Herrmann, S.~Litsey, J.~Stankowicz and J.~Trnka,
  \emph{{\textit{Logarithmic Singularities and Maximally Supersymmetric
  Amplitudes}}}, \href{https://doi.org/10.1007/JHEP06(2015)202}{\emph{JHEP}
  {\bfseries 06} (2015) 202} [\href{https://arxiv.org/abs/1412.8584}{{\ttfamily
  arXiv:1412.8584}}].

\bibitem{Bern:2015ple}
Z.~Bern, E.~Herrmann, S.~Litsey, J.~Stankowicz and J.~Trnka,
  \emph{{\textit{Evidence for a Nonplanar Amplituhedron}}},
  \href{https://doi.org/10.1007/JHEP06(2016)098}{\emph{JHEP} {\bfseries 06}
  (2016) 098} [\href{https://arxiv.org/abs/1512.08591}{{\ttfamily
  arXiv:1512.08591}}].

\bibitem{Bern:2018oao}
Z.~Bern, M.~Enciso, C.H.~Shen and M.~Zeng, \emph{{\textit{Dual Conformal
  Structure Beyond the Planar Limit}}},
  \href{https://doi.org/10.1103/PhysRevLett.121.121603}{\emph{Phys. Rev. Lett.}
  {\bfseries 121} (2018) 121603}
  [\href{https://arxiv.org/abs/1806.06509}{{\ttfamily arXiv:1806.06509}}].

\bibitem{Arkani-Hamed:2014via}
N.~Arkani-Hamed, J.L.~Bourjaily, F.~Cachazo and J.~Trnka,
  \emph{{\textit{Singularity Structure of Maximally Supersymmetric Scattering
  Amplitudes}}},
  \href{https://doi.org/10.1103/PhysRevLett.113.261603}{\emph{Phys. Rev. Lett.}
  {\bfseries 113} (2014) 261603}
  [\href{https://arxiv.org/abs/1410.0354}{{\ttfamily arXiv:1410.0354}}].

\bibitem{Ferro:2016zmx}
L.~Ferro, T.~Lukowski, A.~Orta and M.~Parisi, \emph{{\textit{Yangian symmetry
  for the tree amplituhedron}}},
  \href{https://doi.org/10.1088/1751-8121/aa7594}{\emph{J. Phys.} {\bfseries
  \textbf{A50}} (2017) 294005}
  [\href{https://arxiv.org/abs/1612.04378}{{\ttfamily arXiv:1612.04378}}].

\bibitem{Ferro:2018vpf}
L.~Ferro, T.~Lukowski and M.~Parisi, \emph{{\textit{Amplituhedron meets
  Jeffrey-Kirwan Residue}}},
  \href{https://doi.org/10.1088/1751-8121/aaf3c3}{\emph{J. Phys.} {\bfseries
  \textbf{A52}} (2019) 045201}
  [\href{https://arxiv.org/abs/1805.01301}{{\ttfamily arXiv:1805.01301}}].

\bibitem{Enciso:2014cta}
M.~Enciso, \emph{{\textit{Volumes of Polytopes Without Triangulations}}},
  \href{https://doi.org/10.1007/JHEP10(2017)071}{\emph{JHEP} {\bfseries 10}
  (2017) 071} [\href{https://arxiv.org/abs/1408.0932}{{\ttfamily
  arXiv:1408.0932}}].

\bibitem{Enciso:2016cif}
M.~Enciso, \emph{{\textit{Logarithms and Volumes of Polytopes}}},
  \href{https://doi.org/10.1007/JHEP04(2018)016}{\emph{JHEP} {\bfseries 04}
  (2018) 016} [\href{https://arxiv.org/abs/1612.07370}{{\ttfamily
  arXiv:1612.07370}}].

\bibitem{Karp:2017ouj}
S.N.~Karp, L.K.~Williams and Y.X.~Zhang, \emph{{\textit{Decompositions of
  amplituhedra}}},  \href{https://arxiv.org/abs/1708.09525}{{\ttfamily
  arXiv:1708.09525}}.

\bibitem{Karp:2016uax}
S.N.~Karp and L.K.~Williams, \emph{{\textit{The $m=1$ amplituhedron and cyclic
  hyperplane arrangements}}},
  \href{https://arxiv.org/abs/1608.08288}{{\ttfamily arXiv:1608.08288}}.

\bibitem{Bai:2014cna}
Y.~Bai and S.~He, \emph{{\textit{The Amplituhedron from Momentum Twistor
  Diagrams}}}, \href{https://doi.org/10.1007/JHEP02(2015)065}{\emph{JHEP}
  {\bfseries 02} (2015) 065} [\href{https://arxiv.org/abs/1408.2459}{{\ttfamily
  arXiv:1408.2459}}].

\bibitem{Bai:2015qoa}
Y.~Bai, S.~He and T.~Lam, \emph{{\textit{The Amplituhedron and the One-loop
  Grassmannian Measure}}},
  \href{https://doi.org/10.1007/JHEP01(2016)112}{\emph{JHEP} {\bfseries 01}
  (2016) 112} [\href{https://arxiv.org/abs/1510.03553}{{\ttfamily
  arXiv:1510.03553}}].

\bibitem{IntoTheAmplituhedron}
N.~Arkani-Hamed and J.~Trnka, \emph{{Into the Amplituhedron}},
  \href{https://doi.org/10.1007/JHEP12(2014)182}{\emph{JHEP} {\bfseries 12}
  (2014) 182} [\href{https://arxiv.org/abs/1312.7878}{{\ttfamily
  arXiv:1312.7878}}].

\bibitem{Anatomy}
S.~Franco, D.~Galloni, A.~Mariotti and J.~Trnka, \emph{{\textit{Anatomy of the
  Amplituhedron}}}, \href{https://doi.org/10.1007/JHEP03(2015)128}{\emph{JHEP}
  {\bfseries \textbf{03}} (2015) 128}
  [\href{https://arxiv.org/abs/1408.3410}{{\ttfamily arXiv:1408.3410}}].

\bibitem{4pt1}
J.~Rao, \emph{{\textit{4-particle Amplituhedron at 3-loop and its Mondrian
  Diagrammatic Implication}}},
  \href{https://doi.org/10.1007/JHEP06(2018)038}{\emph{JHEP} {\bfseries
  \textbf{06}} (2018) 038} [\href{https://arxiv.org/abs/1712.09990}{{\ttfamily
  arXiv:1712.09990}}].

\bibitem{4pt2}
Y.~An, Y.~Li, Z.~Li and J.~Rao, \emph{{\textit{All-loop Mondrian Diagrammatics
  and 4-particle Amplituhedron}}},
  \href{https://doi.org/10.1007/JHEP06(2018)023}{\emph{JHEP} {\bfseries
  \textbf{06}} (2018) 023} [\href{https://arxiv.org/abs/1712.09994}{{\ttfamily
  arXiv:1712.09994}}].

\bibitem{4pt3}
J.~Rao, \emph{{\textit{4-particle Amplituhedronics for 3-5 loops}}},
  \href{https://arxiv.org/abs/1806.01765}{{\ttfamily arXiv:1806.01765}}.

\bibitem{Arkani-Hamed:2017vfh}
N.~Arkani-Hamed, H.~Thomas and J.~Trnka, \emph{{Unwinding the Amplituhedron in
  Binary}}, \href{https://doi.org/10.1007/JHEP01(2018)016}{\emph{JHEP}
  {\bfseries 01} (2018) 016}
  [\href{https://arxiv.org/abs/1704.05069}{{\ttfamily arXiv:1704.05069}}].

\bibitem{notesdiff}
S.~He and C.~Zhang, \emph{{\textit{Notes on Scattering Amplitudes as
  Differential Forms}}},
  \href{https://doi.org/10.1007/JHEP10(2018)054}{\emph{JHEP} {\bfseries
  \textbf{10}} (2018) 054} [\href{https://arxiv.org/abs/1807.11051}{{\ttfamily
  arXiv:1807.11051}}].

\bibitem{associahedron}
N.~Arkani-Hamed, Y.~Bai, S.~He and G.~Yan, \emph{{\textit{Scattering Forms and
  the Positive Geometry of Kinematics, Color and the Worldsheet}}},
  \href{https://doi.org/10.1007/JHEP05(2018)096}{\emph{JHEP} {\bfseries
  \textbf{05}} (2018) 096} [\href{https://arxiv.org/abs/1711.09102}{{\ttfamily
  arXiv:1711.09102}}].

\bibitem{halohedron1}
G.~Salvatori, \emph{{\textit{1-loop Amplitudes from the Halohedron}}},
  \href{https://arxiv.org/abs/1806.01842}{{\ttfamily arXiv:1806.01842}}.

\bibitem{Arkani-Hamed:2010gh}
N.~Arkani-Hamed, J.L.~Bourjaily, F.~Cachazo and J.~Trnka, \emph{{Local
  Integrals for Planar Scattering Amplitudes}},
  \href{https://doi.org/10.1007/JHEP06(2012)125}{\emph{JHEP} {\bfseries 06}
  (2012) 125} [\href{https://arxiv.org/abs/1012.6032}{{\ttfamily
  arXiv:1012.6032}}].

\bibitem{Bourjaily:2011hi}
J.L.~Bourjaily, A.~DiRe, A.~Shaikh, M.~Spradlin and A.~Volovich,
  \emph{{\textit{The Soft-Collinear Bootstrap: N=4 Yang-Mills Amplitudes at Six
  and Seven Loops}}},
  \href{https://doi.org/10.1007/JHEP03(2012)032}{\emph{JHEP} {\bfseries 03}
  (2012) 032} [\href{https://arxiv.org/abs/1112.6432}{{\ttfamily
  arXiv:1112.6432}}].

\bibitem{Bourjaily:2015bpz}
J.L.~Bourjaily, P.~Heslop and V.V.~Tran, \emph{{\textit{Perturbation Theory at
  Eight Loops: Novel Structures and the Breakdown of Manifest Conformality in
  $N=4$ Supersymmetric Yang-Mills Theory}}},
  \href{https://doi.org/10.1103/PhysRevLett.116.191602}{\emph{Phys. Rev. Lett.}
  {\bfseries 116} (2016) 191602}
  [\href{https://arxiv.org/abs/1512.07912}{{\ttfamily arXiv:1512.07912}}].

\bibitem{bourjaily2016}
J.L.~Bourjaily, P.~Heslop and V.V.~Tran, \emph{{Amplitudes and Correlators to
  Ten Loops Using Simple, Graphical Bootstraps}},
  \href{https://doi.org/10.1007/JHEP11(2016)125}{\emph{JHEP} {\bfseries 11}
  (2016) 125} [\href{https://arxiv.org/abs/1609.00007}{{\ttfamily
  arXiv:1609.00007}}].

\bibitem{polytopes}
N.~Arkani-Hamed, J.L.~Bourjaily, F.~Cachazo, A.~Hodges and J.~Trnka, \emph{{A
  Note on Polytopes for Scattering Amplitudes}},
  \href{https://doi.org/10.1007/JHEP04(2012)081}{\emph{JHEP} {\bfseries 04}
  (2012) 081} [\href{https://arxiv.org/abs/1012.6030}{{\ttfamily
  arXiv:1012.6030}}].

\bibitem{ArkaniHamed2015}
N.~Arkani-Hamed, A.~Hodges and J.~Trnka, \emph{{Positive Amplitudes In The
  Amplituhedron}}, \href{https://doi.org/10.1007/JHEP08(2015)030}{\emph{JHEP}
  {\bfseries 08} (2015) 030} [\href{https://arxiv.org/abs/1412.8478}{{\ttfamily
  arXiv:1412.8478}}].

\end{thebibliography}\endgroup

\end{document}